\documentclass[a4paper,fleqn]{cas-dc}

\usepackage[numbers]{natbib}
\usepackage{enumitem}
\usepackage{tabularx} 
\usepackage{booktabs}
\usepackage{placeins}
\usepackage{pifont}
\usepackage{amssymb}
\usepackage[table]{xcolor}

\definecolor{principleone}{HTML}{EAF4FF}   
\definecolor{principletwo}{HTML}{F1EAFE}   
\definecolor{principlethree}{HTML}{EAF8EE} 

\usepackage{xcolor}

\newcommand{\kezia}[1]{\textcolor{blue}{\textbf{Kezia says:} #1}}

\begin{document}
\let\WriteBookmarks\relax
\def\floatpagepagefraction{1}
\def\textpagefraction{.001}
\shorttitle{REConnect: Participatory RE for Social Sustainability}
\shortauthors{Damian et~al.}

\title [mode = title]{REConnect: Participatory RE for Social Sustainability}                      
\author[1]{Daniela Damian}[                    
                        ]
                        
\ead{danielad@uvic.ca}
\credit{Provided vision of the Study, Enabled projects by directing the INSPIRE program, Description of REConnect Principles}

\author[1]{Bachan Ghimire}[
                        ]
\cormark[1]                        
\ead{bachan48@uvic.ca}
\credit{Writing of Methodology, Project Descriptions and REActions; Participatory Field Immersion}

\author[2]{Ze Shi Li}[
                        ]              
\ead{zeshili@ou.edu}
\fnmark[1]
\credit{Writing of Generative AI discussion \& Related Work; Provided Expert Knowledge on RE}

\author[1]{Kezia Devathasan}[                        
                        ]
\ead{keziadevathasan@uvic.ca}
\credit{Conducted Interviews and Analyzed Transcripts to Support Findings}

\affiliation[1]{organization={Department of Computer Science, University of Victoria},
                state={British Columbia},
                country={Canada}}

\affiliation[2]{organization={University of Oklahoma}, 
                city={Oklahoma},
                country={USA}}

\cortext[cor1]{Corresponding author}
\fntext[1]{Work completed while conducting PhD at the University of Victoria.}

\begin{abstract}
\noindent \textbf{Context:} Software increasingly shapes daily life, making requirements engineering (RE) essential for ensuring systems contribute to community social sustainability. Yet automated elicitation practices risk distancing RE from the cultural, social, and political contexts that inform user needs, systematically excluding the communities most dependent on socially impactful software. AI-assisted RE has intensified this trend. \\
\textbf{Objective:} This paper introduces REConnect, a human-centered participatory RE framework that re-centers requirements work on human connection and relationality as the foundation for understanding lived experiences and ensuring alignment with community values and aspirations. \\
\textbf{Methods:} REConnect was derived through qualitative analysis of 26 community-engaged software projects conducted through the INSPIRE program at the University of Victoria between 2022 and 2025, spanning rural Nepal, urban Canada, and remote Arctic Canada. We conducted a reflective synthesis across all 26 projects, followed by in-depth thematic analysis of three illustrative projects: BloodSync, a blood coordination platform in rural Nepal; Herluma, a shelter navigation tool for women facing homelessness in Canada; and BridgingRoots, a language revitalization tool for an Indigenous community. Member checking with four community partner representatives established credibility of the derived principles.\\
\textbf{Results:} Three core principles are articulated: building trusting relationships, co-creating with and alongside stakeholders, and empowering users as agents of change. Each is operationalized through actionable REConnect Actions (REActions) embedding relationality and continuous stakeholder engagement throughout the project lifecycle. Member checking confirmed the principles resonate with communities' own accounts of what enabled successful requirements work and sustained system adoption. \\
\textbf{Conclusions:} REConnect positions human connection as the foundation of RE for socio-technical systems aiming toward social sustainability. While AI can accelerate certain RE activities, its integration must be governed by participatory principles that preserve human agency and ensure marginalized voices are not excluded. We discuss how REConnect integrates with AI support while maintaining critical human agency in requirements engineering.
\end{abstract}



\begin{keywords}
Requirements Engineering 
\sep Participatory Design  
\sep Social Sustainability
\sep Human-Centered Design
\sep Generative AI
\end{keywords}

\maketitle

\section{Introduction}

Software is deeply embedded into the infrastructure of modern society, shaping healthcare systems \cite{committee2012health}, transportation networks \cite{national2020reducing}, educational tools \cite{zou2025digital} and our social institutions \cite{werthner2020vienna}. As the boundaries between digital, physical, and social spaces blur \cite{bennaceur2024responsible}, the design of software intensive systems has become inseparable from considerations of \emph{social sustainability}, understood here as the capacity of communities to maintain and improve their well-being over time \cite{ajmal2018conceptualizing}. Software influences the resources that people can access \cite{noble2018algorithms}, the decisions they make \cite{floridi2015onlife}, and even the identities that they form \cite{floridi2015onlife}. Additionally, software systems encode assumptions about  whose needs are prioritized, and whose voices are represented \cite{dodge2011code}. Therefore, the  decisions organizations make in designing these socio-technical systems carry immense ethical, legal, and societal weight, bearing direct consequences for the social sustainability of the communities they affect \cite{moises2023social}. The challenge then, as noted by researchers \cite{Shahin_2022, thew2018value, hussain2022can} and movements such as digital humanism \cite{werthner2020vienna}, is that human values and needs must be placed at the heart of technology development, and must not be treated as secondary concerns. 

Requirements Engineering (RE), as the field pertinent to the elicitation, representation, and validation of requirements for software applications \cite{sommerville1997requirements}, occupies a pivotal position in this challenge. After all, it is during requirements activities that foundational choices are made about whose values, experiences, and well-being are considered in the software design \cite{levina2024incorporating}. More recently, human values like privacy, fairness, and transparency have become critical to system adoption \cite{whittle2019case}, and pure technical functionality is no longer seen as the sole measure of successful software. Yet, achieving socially sustainable software demands more than acknowledging human needs abstractly. User values and requirements are shaped by complex socio-economic and political contexts \cite{le2015strangers}. Additional factors like trust, relationships and power structures influence software adoption in communities \cite{graf2018social}. Recent participatory approaches involving diverse stakeholders sought to identify and operationalize human values in software development \cite{Shahin_2022}, however, there are outstanding challenges in eliciting values, as they are highly contextual and subjective \cite{mougouei2018operationalizing}, perhaps making them easier to understand only \textit{after} software has been deployed. Without sustained attention to these dynamics, we risk building software that is misaligned with human values and needs, undermining social sustainability. 

The increasing scope and scale of today's software brings additional challenges to incorporating human values into requirements. Deciphering users' lived experiences and needs is no longer possible solely through traditional participatory design methods \cite{macaulay2012requirements}. The field of CrowdRE has thus responded with semi-automated techniques such as mining app reviews \cite{groen2017crowd}, sentiment analysis \cite{guzman2014sentiment}, and clustering user feedback with large language models \cite{maalej2025automated} to process user input. While participatory in intention, such methods favor efficiency at the cost of direct user-engagement, risking detachment from the lived experiences that socially sustainable software design needs. These limitations are exacerbated as Generative AI (GenAI) becomes further embedded in RE practice. AI-assisted elicitation tools inherit the biases and gaps of the data they are trained on \cite{mitchell2025fully}, and can undermine human judgement that is necessary to surface subjective human values. As AI agents begin to generate requirements autonomously, the gap between what systems are built to do and what communities actually need risks widening. This threatens to worsen existing societal inequities rather than addressing them. 

Thus, we argue that a participatory "counter-weight" is needed urgently if software is to contribute to social sustainability rather than erode it. In this paper, we introduce \textbf{REConnect}, an RE framework that centers on human connection and relationality as the foundation for understanding lived experiences and positive societal impact on engaged communities. We developed REConnect from analysis of 26 software development projects across rural Nepal, urban Canada, and Canada's remote Arctic, involving multi-stakeholder engagement ranging from four months to two to three years. Drawing on value-based requirements engineering \cite{thew2018value} and community-engaging participatory approaches to system design \cite{whittle2014citizen}, REConnect positions the human connection as central to understanding the lived experiences where impact is sought, and advocates for a human-centered participatory RE approach ’that matters’ to the communities and beneficiaries involved, ensuring
alignment with their values and aspirations. 
It goes beyond capturing functional needs to ensuring that principles of digital humanism, such as fairness,
inclusiveness, and human agency, are accounted for in the context of lived experiences, in ongoing engagement with relevant stakeholders. REConnect embeds a deep understanding of the cultural, socio-economic, and political
contexts in which communities operate, recognizing that these factors influence both the needs expressed and the
feasibility of solutions. We draw on 26 community-facing software projects oriented toward social sustainability outcomes, where requirements emerge as social artifacts negotiated through stakeholder relationships and shaped by cultural contexts. From these projects, we derive three core principles in REConnect: (1) 
building trusting relationships with stakeholders; (2) co-creating with 
and alongside users; and (3) empowering users as agents of change. We describe three example projects to illustrate how these principles operated in practice and contributed to the successful achievement of social sustainability outcomes. In REConnect, each 
principle is operationalized through a set of actionable REConnect Actions (REActions). 
REConnect represents a concrete translation of community-based 
participatory principles into a framework, examined and developed in 
low-resource and marginalized community contexts where traditional RE approaches 
are insufficient.

In what follows, we outline related work in the areas of inclusive software engineering, user-centered and value-based 
requirements engineering, community-based participatory research, and automated, 
AI-assisted RE. We then describe three project examples before 
presenting REConnect, its principles, and associated REActions. We conclude 
by reflecting on the role of sustained human engagement in the era of 
Generative AI, and how REConnect positions humans as guardians of values 
and ethics within AI-assisted RE processes.

\section{Related Work}
\label{sec:relatedwork}

We review four bodies of work that together frame the gap REConnect 
addresses: inclusiveness and human values in software engineering, 
participatory and value-based RE, community-based participatory research, 
and automated approaches to RE. Together they reveal a persistent gap: no 
existing approach addresses the relational, cultural, and trust-dependent 
conditions required for software to contribute to social sustainability in 
low-resource and marginalized community contexts.

\subsection{Inclusiveness in Software Engineering}

Today's complex software systems must account for diverse user experiences
and human contexts \cite{autili2025engineering, bennaceur2024responsible}.
Requirements engineering plays a critical role in understanding user needs
and characteristics, yet organizations often struggle to identify \emph{who}
to involve and \emph{how} to capture their perspectives. This gap can have
serious consequences: for example, object-detection systems in self-driving
vehicles exhibit higher error rates for pedestrians with darker skin tones,
illustrating how insufficiently diverse data can lead to real-world harm
\cite{wilson2019predictive}.

Addressing these challenges requires attention to human values
\cite{friedman2013value} and user characteristics such as age
\cite{mannheim2019inclusion}, ability, and socioeconomic status
\cite{burnett2021toward}. Inclusive software is designed to respect,
reflect, and accommodate the diversity of human needs, abilities, and
contexts \cite{nuseibeh2025engineering}, while human values provide the
ethical and social principles that guide its design and development
\cite{ferrario2016values}. Human values such as equity, fairness,
accessibility, privacy, autonomy, and respect define what inclusivity means
in practice. Without explicitly considering these values, software introduces
bias that can unintentionally marginalize certain groups or create barriers
to participation, directly undermining the social sustainability of the
communities it is meant to serve. Concerns about system biases go back to
Friedman and Nissenbaum \cite{friedman1996bias}, who detailed how computer
systems can perpetuate societal prejudice. Later frameworks such as
Value-Sensitive Design \cite{friedman2013value} aimed to systematically
account for stakeholder values and ethics, while participatory approaches
sought to ensure values are grounded in real user needs and context
\cite{tizard2024elicitation, gama2024awareness}.

Despite these efforts, recent work documents a persistent gap: standard
elicitation techniques, both traditional and crowd-based, fail to
adequately represent diverse user populations, particularly those who are
marginalized or lack digital access \cite{tizard2024elicitation}.

\subsection{User-centered and Participatory RE}

The role of RE in responsible software engineering is
not new. What is continually evolving is our understanding of \emph{context}
necessary to understand diverse users' values, requirements, and lived
experiences in a world increasingly dependent on software
\cite{nuseibeh2025engineering}. Since the early 2000s, authors such as
Sutcliffe \cite{sutcliffe2002user} have advocated for RE activities that
focus on understanding how people think, communicate, and make decisions.
Later frameworks accounted for individual and personal goals and the effect
of time and context on personal requirements \cite{sutcliffe2005personal}.
Context emerged as a driver of RE, as user needs could not be fully
understood without their situational context, including physical setting,
workflow, organizational culture, and domain-specific constraints. Building on the premise that systems should be shaped around users' needs
and contexts, Macaulay's work \cite{macaulay2012requirements} advocates for
direct user involvement in elicitation, analysis, and validation through
group-based and interactive techniques such as joint application development, focus groups, and
cooperative prototyping. Collaboration with users not as data sources but as
active participants is also central to Holtzblatt's contextual inquiry
\cite{beyer1998contextual}, which puts the user's environment front and
center through immersive field studies. Whittle and colleagues
\cite{whittle2014citizen} further expand the circle of participation to
include communities, especially underserved or civic groups, in the
co-production of technology, and argue that human values such as fairness,
transparency, and responsibility must be explicitly surfaced and integrated
into requirements and development cycles \cite{whittle2019values}.

Evidence consistently shows that active user participation supports
adoption, usability, and project success \cite{kujala2003user,
kujala2005role, damodaran1996user}. However, participatory design (PD) has
important limitations. Smith and colleagues \cite{smith2018pd} argue that
sustainable social impact requires PD to move beyond isolated workshops
toward long-term engagement. More critically, Harrington and colleagues
\cite{harrington2019deconstructing} demonstrate that PD as typically
instantiated is a privileged activity that neglects the historical,
access-related, and relational challenges of marginalized communities,
precisely those whose social sustainability depends most on software that
reflects their needs. Standard participatory methods, designed for
organizational settings, do not adequately account for the trust, power, and
cultural dynamics present in community-facing software projects.

Ferrario and colleagues \cite{ferrario2014speedplay} have proposed
Speedplay, a software engineering framework that integrates action research,
participatory design, and agile development to address social innovation
projects involving vulnerable communities. While Speedplay demonstrates the
value of combining these traditions at the process level, it does not
provide a structured RE framework with explicit principles and actionable
practices for elicitation, analysis, and validation in such contexts.

\subsection{Value-based RE}

Awareness of human values in software engineering has grown through the
work of Ferrario et al. \cite{ferrario2016values} and Mougouei et al.
\cite{mougouei2018operationalizing}, drawing on Schwartz's framework of
human values \cite{schwartz2012overview}. Whittle argues that traditional
RE often overlooks the values layer, producing systems that may be
functionally sound but misaligned with users' ethical expectations
\cite{whittle2019values}.

Value-based RE (VBRE) frameworks \cite{thew2018value, harbers2015embedding}
emphasize the explicit elicitation of stakeholder values, motivations, and
emotions, and account for process management implications such as
prioritization conflicts, trade-offs, and stakeholder negotiation
\cite{thew2018value}. Critical systems thinking has been proposed alongside
VBRE to integrate ethics, power, and values into early RE activities
\cite{duboc2019critical}, while Perera et al. \cite{perera2021impact}
provide evidence that considering human values early in RE prevents
violations with financial, reputational, and societal consequences.

A key limitation of existing VBRE methods is their scope: developed for
organizational contexts where stakeholders can be formally identified and
scheduled, they do not address the relationship-building prerequisites,
informal community influencers, or sustained post-delivery engagement that
characterize community-facing software projects where social sustainability
outcomes are at stake. REConnect builds directly on the VBRE tradition
while extending it to these underserved contexts.

\subsection{Community-Based Participatory Research}

Community-based participatory research (CBPR) has emerged over three
decades as a rigorous orientation that foregrounds equitable partnership
between researchers and communities \cite{wallerstein2006cbpr}. More than a
set of methods, CBPR is a relational and ethical commitment to colearning,
mutual benefit, and long-term engagement \cite{wallerstein2010cbpr,
collins2018cbpr}. Jagosh and colleagues \cite{jagosh2015realist} demonstrate
that trust is the central mechanism through which sustained CBPR partnerships
produce community outcomes, and that this trust cannot be manufactured
quickly; it accumulates through repeated, reciprocal engagement over time.

CBPR has been applied productively in health informatics, where Unertl and
colleagues \cite{unertl2015cbpr} found that applying CBPR principles
produced greater community relevance and wider impact, but also identified
technology-specific challenges that CBPR frameworks did not address,
including ownership of technology outputs, build technical capacity with
community partners, and negotiate cultural and temporal mismatches between
academic and community timelines.

Despite this rich tradition, CBPR principles have not been systematically
translated into requirements engineering practice. CBPR addresses \emph{how
to conduct research with communities}; it does not specify how to elicit,
analyze, and validate software requirements, map informal community
influencers into a stakeholder model, or derive formal system requirements
from co-design in low-literacy or low-connectivity contexts. This is the
translation gap that REConnect addresses. Drawing on CBPR's relational
foundations and grounding them in the specific activities of RE, REConnect
provides the first RE-specific operationalization of CBPR principles,
illustrated through three community-engaged projects and supported through
member checking with community partner representatives.


\subsection{Automated Techniques for RE: The Industry Landscape}

The last decade has experienced novel automation support to requirements engineering practices with advances in crowd-based requirements engineering (CrowdRE) and the proliferation of GenAI. 
CrowdRE, defined as an 
umbrella term for automated or semi-automated approaches to gather and 
analyze information from a crowd to derive validated user requirements 
\cite{groen2017crowd}, emerged as a strategy 
to address the challenges of scale in modern software projects. Techniques in CrowdRE commonly involved mining user 
feedback from app reviews, social media, forums, and issue trackers 
\cite{groen2017crowd}, or automated classification to extract requirements 
insights: Guzman et al. \cite{guzman2014sentiment} applied sentiment 
analysis to extract user opinions, while Kurtanovi\'{c} and Maalej 
\cite{kurtanovic2017requirements} used machine learning to classify reviews 
into functional and non-functional requirements. 

The rapid developments in GenAI have advanced these techniques for summarization, categorizing, and synthesizing  crowd feedback, enabling faster processing of user needs \cite{ghosh2024exploring, sami2024ai, almeida2024generative, 
ataei2024elicitron, ronanki2023chatgpt, infoworld2025genai}. Recent studies examine its use for producing draft specifications, generating user stories, synthesizing stakeholder input, and supporting requirements analysis \cite{marques2024using}. Overall, GenAI has brought a broader change to software development, a \emph{“shift to the left”} in which upstream activities become more central to development outcomes \cite{wei2024requirements}. This renewed focus is especially important for RE because requirements work depends heavily on language: it is through conversations, interpretations, negotiations, and natural-language artifacts that stakeholders and technical teams coordinate what a system should do and why it matters \cite{hidellaarachchi2021effects}.

This also means that the quality of requirements carries greater downstream consequences in GenAI-enabled development. When requirements are used as prompts, specifications, or source artifacts for generating code, their clarity and completeness directly shape the quality of the generated implementation \cite{wei2024requirements}. In such workflows, requirements are no longer only planning or communication documents; they become operational inputs into software production. As a result, unclear, incomplete, or conflicting requirements can propagate quickly, with GenAI systems potentially reproducing or magnifying these weaknesses across later development activities \cite{wei2024requirements}.


This makes the inclusiveness failures, already documented in traditional and participatory elicitation contexts, compound in GenAI approaches: less vocal stakeholders are systematically underrepresented, and AI models can inadvertently propagate biases present in training data \cite{tronnier2024bias, alkfairy2024ethical,
ferrara2023fairness}. From a value-sensitive perspective, this highlights
the risk of systematic exclusion, where software may neglect underrepresented
groups or misalign with users' values \cite{aizenberg2020designing,
bano2023operationalizing}. Gender minorities and other marginalized
populations may be overlooked when datasets or training procedures are not
inclusive \cite{undp2024gendersensitive}. Crucially, neither CrowdRE nor
GenAI approaches provide any mechanism for surfacing the informal community
influencers, political dynamics, historical mistrust, or trust-dependent
information sharing that determine whether communities will adopt and sustain
technology solutions. This is not merely a technical limitation but a social
sustainability risk: the communities most in need of socially impactful
software are precisely those whose relational infrastructure is invisible to
automated approaches, and whose exclusion compounds existing inequities
rather than addressing them.

\section{Research Methods}
\label{sec:method}

REConnect's three principles were derived through a two-stage qualitative approach applied to the 26 community-engaged software engineering projects addressing social sustainability challenges, carried out through the INSPIRE program between 2022 and 2025 (Section~\ref{sec:inspire}). We first conducted a \textit{reflective synthesis} across all 26 projects to develop a broad understanding of the conditions under which requirements work succeeded in these community contexts. We then conducted an in-depth \textit{thematic analysis}~\cite{braun2006thematic} of three illustrative projects (Section~\ref{sec:cases}) to refine that broad understanding into the three principles that constitute REConnect. Finally, to examine whether the derived principles reflected community partners' experiences, member checking interviews were conducted with community partner representatives, following established guidance on trustworthiness in qualitative inquiry \cite{mckim2023meaningful}.

\subsection{Research Context: The INSPIRE Program} 
\label{sec:inspire} 

The empirical basis for REConnect's principle derivation is the full corpus of 26 community-engaged software engineering projects carried out through the INSPIRE program\footnote{https://inspireuvic.org} at the University of Victoria, Canada, between 2022 and 2025. INSPIRE is a program of research and social impact that engages teams of senior undergraduate and graduate students in applying design-thinking methodologies to address real-world sustainability challenges identified by partnering communities. Each project uses an experiential learning model that combines an intensive curriculum covering software development, project management, and stakeholder engagement, with an agile, iterative process of problem exploration, prototype design and evaluation, and implementation of solutions in the community. Throughout the process, project members are mentored by industry experts and navigate software development conditions, including ambiguous problem contexts and complex stakeholder relationships.

\begin{table*}[t]
\centering
\caption{INSPIRE Program Project Corpus (2022--2025). The 26 social 
sustainability software projects constituting the empirical basis for 
REConnect's principle derivation. Projects marked * are the three 
illustrative examples presented in Section~\ref{sec:cases}.}
\label{tab:all-projects}
\small
\begin{tabular}{p{0.3cm}p{4.6cm}p{1.3cm}p{2.8cm}p{3.2cm}p{1.3cm}}
\hline
\textbf{\#} & \textbf{Project} & \textbf{Country} & 
\textbf{Community Type} & \textbf{Domain} & \textbf{SDG(s)} \\
\hline
\multicolumn{6}{l}{\textit{2022 Cohort}} \\
1  & ClimAct 
   & Canada & Youth / Environmental 
   & Climate action gamification 
   & 4, 13 \\
2  & Carbon Impact of Web Browsing 
   & Canada & General public 
   & Digital sustainability awareness 
   & 12, 13 \\
3  & Resilient Urban Systems \& Habitat (RUSH) 
   & Canada & Environmental non-profit 
   & Ecological community engagement 
   & 11, 15 \\
4  & Victoria Brain Injury Society (Summer) 
   & Canada & Healthcare / Disability 
   & Acquired brain injury support 
   & 3, 10 \\
5  & Greater Victoria Coalition to End Homelessness
   & Canada & Homelessness services 
   & Shelter coordination 
   & 1, 11 \\
6  & Swan Lake Nature Sanctuary 
   & Canada & Environmental non-profit 
   & Ecological monitoring 
   & 15 \\
7  & EDI Helpline 
   & Canada & University campus 
   & Equity, diversity \& inclusion reporting 
   & 10, 16 \\
8  & The Herluma Project*
   & Canada & Women's shelter services 
   & Homelessness navigation 
   & 1, 5, 11 \\
9  & CIFAL $\times$ Claremont 
   & Canada & Secondary school / UN program 
   & UN SDG education 
   & 4, 17 \\
10 & Victoria Brain Injury Society (Fall) 
   & Canada & Healthcare / Disability 
   & Brain injury support services 
   & 3, 10 \\
11 & Community Transformation Visualization 
   & Nepal  & Rural NGO community 
   & Development program visualization 
   & 1, 17 \\
\hline
\multicolumn{6}{l}{\textit{2023 Cohort}} \\
12 & Bridging Roots* 
   & Canada & Indigenous community (Arctic) 
   & Indigenous language revitalization 
   & 4, 10 \\
13 & FireForce 
   & Canada & Regional government 
   & Wildfire detection \& management 
   & 11, 13 \\
14 & RainWise 
   & Canada & Rural / Agricultural 
   & Drought mitigation / water security 
   & 6, 13 \\
15 & BloodSync* 
   & Nepal  & Rural healthcare 
   & Blood coordination platform 
   & 3, 10 \\
16 & SanghSangai 
   & Nepal  & Rural NGO community 
   & Disaster risk / GBV / livelihood 
   & 5, 13 \\
\hline
\multicolumn{6}{l}{\textit{2024 Cohort}} \\
17 & AI for Non-profits 
   & Canada & Non-profit sector 
   & AI-assisted fundraising tools 
   & 10, 17 \\
18 & BridgingRoots in Tuk 
   & Canada & Indigenous community (Arctic) 
   & Indigenous language \& culture 
   & 4, 10 \\
19 & Educational Tools for Neurodiverse Learners 
   & Canada & University / Education 
   & Neurodiverse learning support 
   & 4, 10 \\
20 & Herluma
   & Canada & Women's shelter services 
   & Shelter bed management 
   & 1, 5, 11 \\
21 & Disaster Preparedness \& Response 
   & Nepal  & Rural government / NGO 
   & Disaster relief coordination 
   & 11, 13 \\
22 & Patient Navigator and Privacy 
   & Nepal  & Healthcare / NGO 
   & Patient health information management 
   & 3, 10 \\
\hline
\multicolumn{6}{l}{\textit{2025 Cohort}} \\
23 & AI for Non-Profits 2025 
   & Canada & Non-profit sector 
   & AI-powered workflow automation 
   & 10, 17 \\
24 & Mathin\'{e} 
   & Canada & University / Education 
   & Educational tools for students 
   & 4 \\
25 & Bridging Roots -- WSANEC 
   & Canada & Indigenous community / School 
   & Indigenous language education 
   & 4, 10 \\
26 & Livelihood Tracking Software 
   & Nepal  & Rural NGO (self-help groups) 
   & Community livelihood tracking 
   & 1, 8 \\
\hline
\multicolumn{6}{l}{Canada: 20 projects \quad Nepal: 6 projects \quad 
Total: 26 projects, 105 Students Engaged} \\
\hline
\end{tabular}
\end{table*}

Table~\ref{tab:all-projects} presents the full corpus of 26 projects, spanning social sustainability domains including Indigenous language revitalization, rural healthcare, environmental management, disability support, disaster preparedness, and education, each involving communities whose access to resources, services, or cultural continuity is at stake, both in Canada and Nepal.

\subsection{Reflective Synthesis as a Research Approach}
\label{sec:refsynth}
The contributions of our work emerge from critical reflection, understood as the examination of lived experiences to surface insights \cite{boud2013reflection, fook2011developing, schon2017reflective}. We characterize our approach as a \textit{reflective synthesis:} the iterative reflection of researchers on practice in which they were embedded, in order to draw a coherent account from many experiences \cite{schon2017reflective}. This approach builds on the tradition of critically reflective practice, which Thompson and Thompson \cite{thompson2021reflective} define as bringing together experience, reflection, and critical thinking in order to revise how one approaches their practice. 
Such reflective approaches have been applied in human-computer interaction (HCI) to derive insight from practitioners' accumulated experience across multiple projects, ranging from broad organizing observations to specific actionable guidance \cite{fook2011developing, kerzner2018framework}. For example, Kerzner \textit{et al.} \cite{kerzner2018framework} reflected on their collective experience across 17 creative visualization-opportunities workshops spanning 10 application domains, a process that, like our own two-stage approach, moved from broad observation toward a specific framework of 25 actionable guidelines. In this mode of inquiry, the researcher's positionality in the practice is treated as the source from which insights are drawn, rather as a threat to the validity of the research. This is because the \textit{synthesis} is integrative: it brings together many situated observations into a set of broad organizing themes, which form the starting point from which REConnect's principles are later refined.

Our material for this reflection was our collective experience together with the data generated across the 26 INSPIRE projects, encompassing students' individual and team reflections, and reflections of the community partner experiences. We, the researchers had  sustained involvement with the projects as program coordinators and project mentors, providing guidance on stakeholder management, communication, and technical practice. In our analysis, we first visited this material individually, reflecting on the projects we had been closest to and noting down what, in our experience, distinguished the engagements in which requirements work succeeded from those in which it did not. We reflected by taking into account: 1) our own experiences while engaged with the projects, 2) the students' experiences, and 3) those of our community partners. We then met over a series of discussions to share these individual reflections, compare them across the breadth of the corpus, and discuss where our accounts converged or differed. Through these successive rounds of discussion, we identified a small set of broad, recurring themes describing the factors that contributed to the successful implementation of the projects within the community: 1)  the importance of relationships developed with stakeholders, 2) iterative collaborative sessions for solution design, and 3) leveraging stakeholders’ motivations in the communities. These broad themes were the output of this stage; they were subsequently deepened and refined through in-depth thematic analysis of three illustrative cases, described in Section~\ref{sec:cases}, from which REConnect's three principles were ultimately derived.

\subsection{From Broad Synthesis to Refined Principles}

To ground and refine these themes in concrete project evidence, two of us conducted an in-depth thematic analysis \cite{braun2006thematic} of student reflections, student interviews, and community partner interviews from the three illustrative projects presented in Section~\ref{sec:cases}: BloodSync, Herluma, and BridgingRoots. These projects were purposively selected to represent variation in geographic setting, community type, and primary RE challenge. We independently coded the material from the illustrative projects, then met over a series of discussions to compare and reconcile the developed codes, resolve discrepancies, and iteratively refine the codes into higher-level concepts. Through these discussions, the refined concepts were progressively shaped into the principles that constitute REConnect. Where the reflective synthesis surfaced broad, recurring patterns across the corpus, this closer analysis grounded those patterns in concrete project evidence. A complete codebook containing the codes, final themes, and exemplary quotes can be found alongside our manuscript as supplementary material\footnote{\url{https://tinyurl.com/REConnect-codebook}}. The resulting principles are presented in section~\ref{sec:reconnect}, and the findings of each stage are reported in Section~\ref{sec:cases}.

\subsection{Data Collection}

In our role as program coordinators and mentors across the INSPIRE program throughout the study period, participating directly in project facilitation and community partner engagement, we had direct access to project artifacts, community partner interactions, and team dynamics. This insider positionality is, as described in Section~\ref{sec:refsynth}, the epistemic basis for the reflective synthesis stage: our sustained involvement is what made the patterns underlying REConnect's principles visible in the first place. At the same time, this same closeness to the work requires caution to ensure that the broad themes and principles we report reflect the corpus as a whole rather than our individual interpretation. We address this through independent reflections prior to joint reconciliation in both the reflective synthesis and thematic analysis stages, and through member checking with community partners, described in section~\ref{sec:memberchecking}. 

Our data collection procedures included personal weekly reflections as well as team reflections publicly documented on the INSPIRE program website and project report writing completed at the end of each project cycle. In addition, semi-structured interviews were conducted with community partner representatives and student project members during the course of the projects, providing direct participant voice on requirements processes, stakeholder relationships, and design decisions as they unfolded. 

\subsection{Member Checking}
\label{sec:memberchecking}
To establish the credibility of the derived principles, we conducted member checking interviews with community partner representative from 4 of the 26 INSPIRE projects: \textit{Bridging Roots} and \textit{Herluma} in Canada, and \textit{Community Transformation Visualization} and \textit{Livelihood Tracking Software} projects in Nepal. These four projects were selected to represent variation in geographic setting and community type, and to provide a check on the derived principles that extends beyond the three illustrative cases. Two of the four member checking projects (Bridging Roots, Herluma) are the illustrative cases themselves, while the third and fourth, Livelihood Tracking Software and Community Transformation Visualization were included to examine whether the principles resonated with a community partner's experience outside the cases used to derive them, offering an additional, independent check on generalizability beyond the corpus subset that informed the thematic analysis. Each interview presented the three derived principles and their associated REActions and invited partners to reflect whether and how the principles resonated with their experience of how the partnership and requirements work developed. Partners were explicitly invited to identify anything missing, overstated, or inconsistent with their experience, and responses were analyzed to identify confirmations, qualifications, and contradictions of the derived principles.

\section{REConnect in Practice: Project Examples}
\label{sec:cases}

The exploratory reflective analysis across the 26 projects surfaced three broad, recurring themes in how teams engaged with their community partners: the importance of the relationships developed with stakeholders, the value of iterative collaborative sessions for solution design, and the practice of leveraging stakeholders' motivations within their communities. At this level the themes described the general contours of effective community-partnered work, identifying what teams consistently attended to, however, they do not the underlying mechanisms or the specific quality each theme demanded.

The subsequent in-depth thematic analysis sharpened these contours, with nineteen codes consolidating into the same three themes but with a markedly deeper character. The first theme moved from relationships in general to \textbf{relationships of trust}. Codes such as site \textit{visits to partner communities}, \textit{formal and unmediated CP meetings}, \textit{role-aware communication}, \textit{multi-level stakeholder mapping}, \textit{understanding stakeholder values}, \textit{sustained contact across project phases}, and \textit{cultural and linguistic accommodation} showed that trust was not assumed but built, emerging through repeated, direct, and contextually attuned engagement rather than a single mediated exchange. The second theme moved from iterative sessions to \textbf{co-creation}, a mode of collaboration in which stakeholders act as active participants in shaping the design rather than being merely consulted. Here, codes including \textit{drift correction}, \textit{field-based assumption validation, specification confirmation with partners, revision upon immersion, }and \textit{CP-led artifact review} revealed that design decisions were continually surfaced, challenged, and reworked alongside partners rather than handed to them for sign-off. The third theme moved from leveraging motivations to \textbf{stakeholders being empowered} on the basis of those motivations. Codes such as \textit{surfacing distinct stakeholder concerns, navigating stakeholders' motives, partner-initiated role definition, communication expectations, stakeholder-driven scope change,} and \textit{capacity building and training} showed that partners were given defined roles, genuine influence over project direction, and the means to sustain the work, turning their motivations into agency rather than mere buy-in. These three refined themes, trust, co-design, and empowerment, constitute the \textbf{three REConnect principles} we develop in the remainder of this section. Table \ref{tab:codebook} provides a partial codebook demonstrating how our codes were shaped into the REConnect principles. A full codebook has been provided alongside this manuscript as supplementary material.

The following three projects are drawn from the INSPIRE corpus as 
illustrative examples of REConnect's principles in practice. Selected 
to represent variation across projects, they show how the three principles derived from the reflective synthesis and subsequent thematic analysis manifested 
across structurally distinct community contexts: rural Nepal, urban 
British Columbia, Canada, and remote Tuktoyaktuk in Arctic Canada. In each case, the insights 
that determined whether the system would be adopted, trusted, and 
sustained by its community could only be reached through sustained 
relationships, not through conventional elicitation. That the same 
three principles operated consistently across these structurally 
distinct contexts is itself a meaningful result, further supported by 
their consistent presence across the broader 26-project corpus from 
which they were derived.

\begin{table*}[t]
\centering
\caption{Partial Codebook: three REConnect principles with two example codes and quotes for each.}
\label{tab:codebook}

\begin{tabularx}{\textwidth}{@{}p{3.2cm}X@{}}
\toprule
\textbf{Code} & \textbf{Example quote} \\
\midrule

\rowcolor{principleone}
\multicolumn{2}{@{}l}{\textbf{Principle 1: Building trust through shared presence.}}\\[2pt]

Site visit to partner community
& ``It’s incredible to witness the strength that the community has, with each member wanting to see it succeed and rebound from the effects of colonization and residential schools. They will succeed with or without our help; you have to be hardy to live in such a place. It’s unbelievable to be a part of such a much larger movement.'' \\[4pt]

Role-aware communication
& ``The approach we had to take with the administrative director, the medical director, the representatives from the Red Cross, the mayor and the other bureaucratic leaders; for the same project all those things had to be different. You cannot talk with them in the same way.'' \\[6pt]

\rowcolor{principletwo}
\multicolumn{2}{@{}l}{\textbf{Principle 2: Co-creating with community \& political stakeholders}}\\[2pt]

Drift correction
& ``At the mid of the project we got to spend the day with community partners and work together to understand that reality was different; we understood why our requirements were changing constantly.'' \\[4pt]

Field-based assumption validation
& ``Although we had prepared the system with certain assumptions about the place in mind, some of those assumptions were challenged on the field. Some were verified, those were confirmed.'' \\[6pt]

\rowcolor{principlethree}
\multicolumn{2}{@{}l}{\textbf{Principle 3: Empowering users as agents of change}}\\[2pt]

Surfacing distinct stakeholder concerns
& ``One of my takeaways was how the community speaks about people dealing with substance abuse. There is a large understanding of why and a collective goal to heal as a community, which was really heartwarming to see.'' \\[4pt]

Partner-initiated role definition
& ``With the local leaders, they were like: what should we do for you? What should be our role in the whole project?'' \\

\bottomrule
\end{tabularx}
\end{table*}

\subsection{Project BloodSync}

Nepal's healthcare landscape is characterized by a stark urban-rural divide: more than 70\% of the population cannot afford private care, and remote districts like Rukum lack essential infrastructure \cite{Karkee2016Nepal}. Chaurjahari Hospital Rukum (CHR) faces chronic blood shortages due to the absence of a local transfusion center. The BloodSync project developed an ICT-driven blood management platform tailored to the region's low-literacy, low-connectivity, and socio-cultural realities, with an emphasis on maternal and emergency care.

\textbf{\textit{Building trust through shared presence.}} In remote communities like Rukum, medical care depends more on local relationships than formal systems \cite{Panday2017Health}. Rather than assuming solutions, the team prioritized learning from lived experience on the ground. The BloodSync team journeyed over 600 kilometers across landslides and earthquake-damaged terrain to be physically present in the community, spending time in hospital wards and participating in blood donation alongside staff. Through sustained informal dialogues with hospital staff, village elders, and health volunteers, the team discovered that \emph{trust had to precede design}: cultural fears around blood donation, including beliefs that donation weakens the body or shortens life, rooted in misinformation and historic institutional neglect, would not have surfaced in a short-term requirements elicitation session. The team learned that the community valued humility, integrity, and credibility \cite{kathayat2024investigating, subedi2025ethical}, values that had been notably absent in previous failed development efforts in the region, and that political figures such as ward leaders were \emph{informal credible, influencing stakeholders} whose buy-in was a prerequisite for any solution's legitimacy. Trust with the Mayor, once established, was what unlocked access to ward leaders: local political figures reporting to the Mayor, the ward leaders would be central to awareness campaigns and donor mobilization.  The team carefully mapped the political and cultural dynamics among ward leaders and health institutions. The Mayor emphasized their role directly, and mapping these political and cultural dynamics before any design work began secured early commitment from actors who later became co-design partners, a pattern we would observe consistently across the broader project corpus. 

\textbf{\textit{Co-creating with community and political stakeholders.}} Initial design assumptions were directly challenged through early feasibility discussions. The team had planned a self-registration donor app, but community engagement revealed that \emph{the root problem was not technical but relational}: limited digital literacy, superstition-driven resistance to donation, and the absence of trusted mobilization channels. An alternate design emerged through co-production with hospital staff, ward leaders, and Red Cross representatives. Rather than automated alerts, the system routed blood requests through ward leaders who would contact donors personally, embedding existing trust relationships into the technical workflow. Immersive co-design workshops with these stakeholders surfaced accessibility requirements around language, iconography, and offline use that would not have been specifiable without their participation. \emph{The requirements were not gathered from stakeholders; they were produced with them.} 

\textbf{\textit{Empowering ward leaders as agents of sustained change.}} A critical insight from BloodSync was that \emph{empowerment must be structurally embedded, not added on}. As discussions evolved, it became clear that giving ward leaders an autonomous coordination role aligned naturally with their existing civic responsibilities and intrinsic motivation to serve their communities. Simulation-based training, coordinated by the Mayor, built digital confidence among ward leaders and Red Cross staff who were initially unfamiliar with the tools. Their roles evolved from passive participants to active system owners, culminating in the formation of a formal blood coordination committee. Community leaders subsequently proposed extending the digital workflow to maternal health and emergency transport. The outcome was more than a technical solution: it was a \emph{locally embedded care infrastructure} sustained by the relationships and ownership structures that the RE process had helped to build.

\subsection{Project Herluma} 

Improving access to shelters for women fleeing intimate partner violence and facing homelessness has been the goal of the Herluma project, conducted in collaboration with the Greater Victoria Coalition to End Homelessness (GVCEH) in Victoria, British Columbia. Through a series of design pivots driven entirely by community engagement, the project produced a web-based fleeing-violence navigator that aggregates real-time shelter bed availability for shelter staff. 

\textbf{\textit{Building trust through authentic engagement with lived experience.}} Forging genuine relationships with women experiencing homelessness and with shelter staff required more than technical readiness. The team began by attending sessions led by people with lived homelessness experience, which surfaced the depth of intergenerational trauma and systemic complexity that \emph{no requirements document could have conveyed} \cite{burrows2019motivational, shinn2007international}: specifically, how generational cycles of abuse and poverty shape the conditions that make homelessness not an individual failure but a structural outcome. These sessions were foundational: the team's demonstrated authenticity in understanding the problem at its core was what opened the door to a working relationship with GVCEH. At the shelters, shelter staff were navigating daily emotional tolls of trauma disclosures, safety planning, and hours spent phoning other shelters in search of vacancies; any technology-based intervention risked being dismissed unless it first proved capable of recognizing that burden. With GVCEH's guidance, the team subsequently built credibility with BC Housing, the broader government-owned entity in the Province and whose endorsement signaled to hesitant shelter organizations that this project was different from previous short-term interventions. \emph{Institutional trust was earned sequentially, not assumed simultaneously}: each relationship enabled the next. 

\textbf{\textit{Co-designing across a complete reversal of initial assumptions.}} The team's initial assumption was that women needed a mobile application for locating shelters and accessing resources. Co-design sessions with women with lived experience \emph{invalidated the entire design premise}: most did not own a cell phone, many required anonymity to stay safe from abusers, and a digital footprint posed a direct safety risk. These insights, accessible only through sustained and trusting engagement, led to a complete overhaul of the requirements. The solution pivoted to a web-based tool for shelter operators, incorporating pseudonyms, unique identifiers, and configurable visibility settings to protect shelters serving undocumented or high-risk populations. \emph{The co-design process did not refine a design; it replaced one.} This was a recurring observation across the INSPIRE corpus: co-design's most important contribution is often the requirements it eliminates, not just the ones it surfaces. 

\textbf{\textit{Empowering shelter operators as advocates and owners.}} Most shelter operators were women who had previously experienced homelessness themselves. Their sustained engagement in the design process was not merely instrumental; it was \emph{intrinsically empowering}. Shaping a system that could help women in situations they had lived through, and whose friends and family remained on the streets, gave operators a sense of agency rooted in the desire to give back to their own communities \cite{deci2013intrinsic}. Early adopters like the Victoria Native Friendship Centre integrated the tool into routine operations and became internal advocates, encouraging other shelters to adopt. Staff-led feedback informed key adaptations to privacy controls, reinforcing operators' sense of ownership over how and when information was shared. The project demonstrated that \emph{empowerment and adoption are not separate outcomes}: when stakeholders genuinely own a system, adoption follows.

\subsection{Project BridgingRoots} 

"How do we teach a lost Indigenous language in a remote community in Canada’s Arctic with scarce resources in a way that respects community participatory action?" is a quote by  M. Tomasino, Vice-Principal of Mangilaluk School in Tuktoyaktuk (Tuk), a remote Inuit hamlet of about 950 inhabitants on the Western Canadian shores of the Arctic Ocean. "The impact of residential schools in Canada is that our language and culture were disallowed in schools after 1871, leaving Indigenous people without the means to communicate their stories." The Bridging Roots project co-designed, in an authentic partnership with the Tuktoyaktuk community, a game-based language revitalization tool to teach Inuvialuktun (the language of the Western Canadian Inuit) to students in grades 2-12 at the Mangilaluk School. When using the tool, the students engage in class activities in the form of games that use the Inuvialuktun language; they can also add to the information about their community and land (stories, photos), actively contributing to a growing body of knowledge about their language and community.

\textbf{\textit{Building trust across cultural and historical distance.}} Foundational to a successful interaction with Indigenous communities is respect for their people, culture, and ways of living \cite{wilson-raybould2024true}. The project evolved from the high goal of involving Inuit youth in learning about technology design to a more specific goal of supporting their community in its language revitalization efforts. Over two years, the project team developed a trusting relationship with the community and the Mangilaluk School in Tuk that proved essential to the discovery, refinement, development, and validation of tool features, made possible through sustained engagement with youth and respected Elders. The long-term engagement, three week-long in-person visits followed by ongoing online communication, led to cultural sensitivity and identifying the community's value of authentic relationships and community-participatory action \cite{wilson-raybould2024true}, rooted in historical trauma due to extensive extractive relationships. The interactions with the Inuit youth were informal in the first few months, mostly texting about the youth's life on the shores of the Arctic Ocean and which, on the surface, was not contributing to the project's success. Remarkable, however, are the friendships that developed between the Indigenous youth and the non-Inuit team members: authentic and deep, in line with the Indigenous culturally-informed ways of living and being. The Inuit youth became more engaged in the project with every visit to Tuktoyaktuk, leading to meaningful co-design of the language teaching tool.

\textbf{\textit{Co-designing with a community whose knowledge systems shaped the requirements}} was essential to understanding the intended impact for language revitalization and to developing the system features. In-person workshops during immersive visits to Tuk engaged community Elders, language teachers, and students across grades 2 through 12, revealing the project team's limited understanding of the Indigenous ways of teaching, deeply embedded in ways of living on traditional lands. Indigenous languages carry meaning rooted in cultural traditions and ways of interacting with the environment: community consultations revealed that in oral history cultures like the Inuit's, traditions are transmitted through stories, informing features such as recording Elders' accounts of fishing, hunting, and life on the land. Games emerged as the most effective engagement format given that low attendance and depression, typical of Indigenous communities where students manage significant family-related trauma, are rarely considered in educational tool design. Co-design workshops also surfaced conflicting community values between timely language teaching and correctness of pronunciation, a tension made urgent by only four of approximately 100 families in Tuk still speaking Inuvialuktun fluently; the collective resolution was to enable language teachers to verify all content in the app.

\textbf{\textit{Empowering youth and community as knowledge keepers}} in creating and growing the language content has been a key success factor in the adoption, uptake, and sustainability of the tool in the classrooms. Motivated by a desire to learn tech design, the Inuit youth used Figma to design games, often in collaboration with grade two students who drew game elements that would appeal to the younger students. The sense of engagement further motivated the older students to interview community Elders and record Inuit stories about traditional skills, including hunting or fishing, or simply surviving on the ice, in a place surrounded by ice for ten months in a year. \emph{The community was not the subject of the design process; it became its author.} New content being taught in the language app was now being created by their own community members! Finally, the youth empowerment extended beyond actively participating in the technology-enabled language revitalization in their community: at the time of writing, three of the Inuit youth, inspired to pursue higher education ``down south", are submitting applications for undergraduate degrees at the University of Victoria. They declared a higher confidence to live outside their close-knit Indigenous community, and about 4,000 km away from their families.

\section{REConnect: Participatory RE for Social Sustainability}
\label{sec:reconnect}

Drawing on our insights in the INSPIRE project, REConnect is a requirements engineering 
framework that embeds users, community members, and other stakeholders 
throughout RE activities during system development and post-delivery in 
pursuit of social sustainability outcomes. In the era of data-driven 
automation techniques for understanding user needs and requirements 
\cite{groen2017crowd}, there is a risk of overlooking the individual behind 
each data point. While stakeholder values such as trust, collaboration, and 
ethics \cite{thew2018value} are shaped by lived experiences in context, and 
are fundamental to software that contributes to community well-being, 
conventional elicitation alone cannot reliably surface them.

As established in Section~\ref{sec:relatedwork}, CBPR principles have 
addressed this challenge in other fields but have not previously been 
operationalized within the RE lifecycle \cite{wallerstein2006cbpr, 
unertl2015cbpr}. REConnect bridges that gap. Drawing on value-based 
requirements engineering \cite{thew2018value} and grounded in CBPR's 
relational and ethical foundations \cite{wallerstein2006cbpr}, REConnect 
grounds solution design in the lived experiences of communities where 
social sustainability is at stake, ensuring alignment with community 
values and aspirations while accounting for the cultural, socio-economic, 
and political contexts that shape both what communities need and what 
solutions are feasible there.
\\\\
\noindent REConnect is organized around three principles:

\begin{itemize}
    \item \textbf{Building trusting relationships with stakeholders}, which enables the
    identification of stakeholder roles, including formal and informal
    influencers, and the values, motivations, and emotions that shape
    their participation
    \item \textbf{Co-creating with and alongside users}, where key
    stakeholders contribute to shaping requirements and design from their
    lived experience, not merely as feedback providers
    \item \textbf{Empowering users as agents of change}, shifting their
    role from passive recipients of design decisions to active shapers
    of outcomes
\end{itemize}

Each principle is described below with examples from the three case
studies. To make REConnect actionable, we distilled recurring practices
observed across the broader project corpus into a set of REConnect
Actions (REActions).

\subsection{Building Trusting Relationships with Stakeholders}

Trust is the foundation of effective requirements engineering for 
socially sustainable software, and it cannot be manufactured quickly. 
Requirements are negotiated, socially-constructed artifacts 
\cite{potts1997naturalistic}, and the willingness of stakeholders to 
share sensitive information, including cultural dynamics, political 
constraints, and personal values, depends on whether they believe their 
concerns will be heard and acted upon \cite{burnay2017trust}. Yet 
existing RE methods have largely treated trust as an assumed 
precondition or a transactional variable in elicitation, rather than as 
a foundational construct that must be actively cultivated across the 
full RE lifecycle. Burnay and Snoeck \cite{burnay2017trust} provide an 
empirical study of trust specifically in RE elicitation; REConnect 
extends this work by grounding trust as a longitudinal, cross-cultural 
construct derived from community-facing software projects spanning 
diverse low-resource and marginalized contexts.

Across the INSPIRE corpus, trust consistently emerged as a prerequisite 
for substantive requirements discovery, not a parallel activity or a 
soft complement to technical elicitation. 
For example, in the Victoria Brain Injury Society project, students worked on designing simplified web resources for people with acquired brain injuries. The goal was to make important information easier to find, understand, and use, especially for individuals who may experience challenges with memory, attention, or cognitive load. The community partner from the project emphasized that building trust before formal project activities was especially important when working with vulnerable community members: \textit{"I think it really helped that the students got to know some of the patients first, before interviewing them and getting feedback on the prototypes. Some of our patients are especially vulnerable and it helps them to have a sense of familiarity, and to feel more comfortable with the students first." (Project \#10, Community Partner)}.  

In BloodSync, humility and 
credibility demonstrated over extended field visits unlocked politically 
motivated constraints and culturally rooted barriers that would never 
have surfaced in a structured elicitation session. As the community 
partner noted: \textit{"If the students did not have a prior connection with us (the hospital), they would have never been allowed inside the Mayor's office. They seemed like hard working students who had a good reason and seemed to understand our problems here. That's why I could ask our director to schedule a meeting with the Mayor to discuss software development"(Project \#15, Community Partner)}. 

The Herluma project similarly highlighted the importance of trust-building when working with community members who had experienced significant personal hardship. From the community partner’s perspective, some participants were understandably hesitant to engage with students at first:
\textit{“I would say it took some convincing to have some of them agree to talk to the students. What they have gone through is unimaginable and naturally they are a bit hesitant at first.” (Project \#8, Community Partner)}

Students also recognized this hesitation and described how participants became more open once relationships had begun to form: \textit{“I feel like it took a lot of trust-building to have the focus group with the women. They seemed very hesitant at first, but once they got to know us a bit more, I felt like they were more open about their experiences and helped us with our project.”(Project \#8, Student)}

In each project, \emph{trust 
preceded requirements}: without it, the most important design 
constraints remained invisible and social sustainability outcomes 
unachievable.

\vspace{1em}
\noindent\textbf{REConnect Actions for Building Trusting Relationships:}
\begin{itemize}
    \item \emph{Contextual Stakeholder Mapping}: Identify stakeholders that play formal and informal roles as community influencers (e.g., ward leaders in 
    BloodSync, Elders in BridgingRoots). Together, co-create stakeholder maps to 
    surface relational dynamics and their implications for the system 
    \cite{alidoosti2023stakeholder}, including the cultural and 
    political landscape that shapes whose voice carries weight and whose 
    needs are at risk of being overlooked.

    \item \emph{Immersive Field Presence}: Commit to sustained field 
    visits and cultural immersion to understand context that cannot be 
    conveyed remotely \cite{beyer1999contextual}. Use ethnographic 
    methods such as storytelling sessions and informal focus groups to 
    build empathy and relational credibility with community members 
    before elicitation begins.

    \item \emph{Community-Engaging Facilitated Workshops}: Co-host 
    problem-framing workshops \cite{simonsen2013routledge} with local 
    organizations who understand the community context. Conduct 
    story-sharing sessions to surface historical mistrust, cultural 
    beliefs, and systemic barriers that shape both what communities need 
    and what solutions are feasible for sustaining community well-being.
\end{itemize}

\subsection{Co-Creating With and Alongside Users}

Participatory design and value-based RE have both long advocated for
stakeholder involvement in software development \cite{smith2018pd,
thew2018value}. What remains underspecified, however, is how co-creation
integrates with the formal structural activities of RE: requirement
decomposition, scenario simulation, constraint negotiation, and artifact
validation, particularly in contexts where communities have been
historically excluded from shaping the technologies that affect them
\cite{harrington2019deconstructing, tizard2024elicitation}. REConnect
addresses this gap by treating co-creation not as a consultation layer
added to RE but as the mechanism through which formal requirements are
produced, challenged, and legitimized in pursuit of socially sustainable
outcomes.

Across the INSPIRE corpus, the most significant requirements shifts
occurred not through elicitation but through sustained collaborative
engagement: co-creation exposed not only what communities needed but what
the design team had fundamentally misunderstood. In Bridging Roots, Workshops with Elders, teachers, and students revealed that the team’s understanding of language learning was rooted in a Western educational framework that did not apply. The community’s oral traditions, land-based knowledge, and ways of transmitting culture became direct functional requirements. As one community partner reflected, \textit{"Just getting the kids thinking about the language is good… but some of it is not just their age, but the reality of life for many Indigenous people now. They just feel kind of almost embarrassed sometimes to feel like they’re putting themselves out there." (Project \#12, Community Partner)}. 
This perspective directly reinforced why co-creating with and alongside users was essential in BridgingRoots. Rather than imposing a Western model of classroom participation, the team had to listen closely to how community members actually experienced language learning.

Similarly, for the ClimAct project, students designed an app for youth that gamified taking more eco-friendly actions. This project highlighted the importance of co-creation, especially when building tools for young users. Rather than creating only for youth, students involved them directly in the design process, allowing their ideas, drawings, and perspectives to shape the final concept. As the community partner explained: \textit{“I loved the way the team came in and designed things with the kids. It was great for their learning, and I think it made them feel like they were really a part of the process..I know some of them really enjoyed it and actually put quite a lot of thought into their drawings and ideas for you guys.”(Project \#1, Community Partner)}
    
In all projects, \emph{the requirements were not gathered from
stakeholders; they were produced with them.} Co-creation's most important
contribution is often the requirements it eliminates as much as the ones
it surfaces, a finding consistent with the broader INSPIRE corpus and
not captured by existing co-design frameworks operating outside the RE
lifecycle \cite{ferrario2014speedplay}. Where communities have been
historically excluded from shaping the technologies that affect their
social sustainability, this distinction is not merely methodological
but consequential.

\vspace{0.5em}
\noindent\textbf{REConnect Actions for Co-creating Alongside Users:}
\begin{itemize}
    \item \emph{Iterative Prototyping and Simulation}: Use role-playing
    scenarios and simulated workflows to surface usability gaps with
    diverse stakeholders (e.g., emergency scenario simulations in
    BloodSync). Employ Wizard of Oz prototypes
    \cite{segura2013uiskei++} for rapid feedback on incomplete
    features.

    \item \emph{Community-Engaged Requirement Mapping}: Conduct joint
    workshops to decompose requirements with explicit attention to
    cultural, economic, and technical constraints. Use community
    feedback loops to validate and challenge design assumptions (e.g.,
    the shift from SMS broadcasts to a multi-tier coordination workflow
    in BloodSync).

    \item \emph{Co-Design Artifacts Within Socio-Economic Context}:
    Develop contextual artifacts, including paper sketches, mockups,
    and dashboards, suited to the literacy levels and environments of
    the users (e.g., icon-based designs for BloodSync, desktop-first
    prototypes for Herluma). Iterate these artifacts with users until
    they reflect local realities rather than designer assumptions.
\end{itemize}

\subsection{Empowering Users as Agents of Change}

Existing RE methods do not treat post-delivery community ownership as
a formal RE outcome. Empowerment appears in participatory design and
CBPR literature as a goal of community engagement
\cite{wallerstein2006cbpr, collins2018cbpr}, and in information systems
research as a predictor of technology adoption \cite{deci2013intrinsic,
ryan2000self}, but neither tradition has operationalized it within the
RE lifecycle as a traceable consequence of how requirements work is
conducted. REConnect makes this claim explicitly: across the INSPIRE
corpus, sustainable adoption was consistently achieved in projects where
stakeholders transitioned from passive beneficiaries to active owners
with recognized roles in the system's ongoing governance, and this
transition was a direct product of how the RE process was structured,
not an outcome of the system itself. Where social sustainability depends
on communities continuing to use, adapt, and govern software beyond the
project team's involvement, this distinction is foundational.

In BloodSync, inviting ward leaders into mayor-led training and giving
them autonomous coordination roles transformed the blood donation
workflow into a community-governed practice, with a formal committee
continuing to refine and expand it after the project ended. The community
partner reflected on this shift: \textit{"Now that the ward leaders have been trained to receive requests from hospital, they will reach out to their respective wards members and directly ask if they are available for blood donations."(Project \#15, Community Partner)}

The Climact project also showed how co-design can be empowering for young users. By seeing their own drawings, ideas, and suggestions reflected in the final app, the students were not only more likely to engage with the tool, but also able to recognize that their perspectives had meaningfully shaped the design. As the community partner reflected: \textit{"I do think that the students are more likely to use the app because they see themselves in the design of it. I think that must be empowering for them to see that these university students literally built their ideas into an app, I imagine that’s quite something for a seventh grader." (Project \#1, Community Partner)}

Similarly, in Livelihood Tracking Software project, a member of a rural Nepal women's self-help group noted that the software  allowed groups to learn from each other's success factors and was empowering to them. She notes: \textit{"We can look at our neighbor groups and learn from their investment in crops and livestock and boost our own sales" (Project \#26, Community Partner)}

In each project, \emph{empowerment was not a byproduct of the design};
it was a direct consequence embracing community values with trusting relationships and co-created solutions. When communities are empowered through trust as an RE process 
itself, the result is software that communities sustain because they 
genuinely own it, a condition for social sustainability that no 
technical solution alone can achieve. The relationship between RE 
practice and post-delivery community ownership is examined 
comparatively in Section~\ref{sec:discussion}.

\vspace{0.5em}
\noindent\textbf{REConnect Actions for Empowering Users as Agents of
Change:}
\begin{itemize}
    \item \emph{Motivation Mapping}: Explicitly analyze why stakeholders
    care and how the system can align with their intrinsic motivations
    (e.g., ward leaders motivated by civic accountability; Inuit youth
    and teachers driven by cultural preservation). Map these motivations
    to specific system capabilities that reinforce community agency and
    long-term ownership.

    \item \emph{Engage Empowered Stakeholders}: Involve stakeholders
    directly in designing the capabilities that correspond to their
    roles and motivations (e.g., youth co-designing game workflows in
    BridgingRoots; shelter workers shaping visibility and anonymity
    features in Herluma).

    \item \emph{Capacity Building and Training}: Embed training within
    co-design activities so stakeholders gain the skills to own and
    sustain the system beyond the project team's involvement (e.g.,
    mayor-led digital training for ward leaders in BloodSync),
    ensuring that empowerment outlasts the project itself.
\end{itemize}

Taken together, the three principles form an interdependent cycle rather
than a sequential checklist. Trust creates the relational conditions under
which co-creation becomes substantive; co-creation surfaces the motivations
that make empowerment possible; and empowerment sustains the community
ownership that keeps requirements relevant beyond delivery. 
Table~\ref{tab:REConnect} summarizes the REConnect principles and their
associated REActions. Table~\ref{tab:case-study-comparison} synthesizes
how each principle manifested across the three projects and the
distinctive insight each project contributed to the framework's derivation.

\begin{table}[H!]
\centering
\caption{REConnect Principles and their Associated REActions}
\label{tab:REConnect}
\renewcommand{\arraystretch}{1.4}
\begin{tabularx}{\linewidth}{lX}
\toprule
\textbf{REConnect Principle} & \textbf{Associated REAction} \\
\midrule
\multirow{3}{*}{\parbox{3cm}{\textbf{Building Trusting\\Relationships}}}
  & Contextual Stakeholder Mapping \\
  & Immersive Field Presence \\
  & Community-engaging Workshops \\
\midrule
\multirow{3}{*}{\parbox{3cm}{\textbf{Co-Creating With\\and Alongside Users}}}
  & Iterative Prototyping \\
  & Requirement Mapping \\
  & Co-Create Artifacts \\
\midrule
\multirow{3}{*}{\parbox{3cm}{\textbf{Empowering Users as\\Agents of Change}}}
  & Motivation Mapping \\
  & Engage Empowered Stakeholders \\
  & Capacity Building and Training \\
\bottomrule
\end{tabularx}
\end{table}

\begin{table*}[t]
\centering
\caption{Cross-project synthesis: how REConnect principles manifested across
the three projects and the distinctive insight each case contributed
to REConnect.}
\label{tab:case-study-comparison}
\small
\renewcommand{\arraystretch}{1.4}
\begin{tabular}{p{1.6cm}p{2.2cm}p{3.0cm}p{3.0cm}p{3.0cm}p{2.4cm}}
\hline
\textbf{Project} &
\textbf{Context \& Primary RE Challenge} &
\textbf{Trust: How it was built} &
\textbf{Co-Creation: Key contribution} &
\textbf{Empowerment: Outcome} &
\textbf{Distinctive insight for REConnect} \\
\hline
\textbf{BloodSync} &
Rural Nepal; navigating political and cultural gatekeeping in a
low-connectivity, low-literacy context &
Physical presence; engagement with Mayor, ward leaders, and Red Cross;
mapping political and cultural dynamics before any design work &
Replaced donor-registration app with ward-leader-mediated notification
model; formal requirements derived from community coordination logic &
Ward leaders took autonomous coordination roles; blood coordination
committee formed; Red Cross staff gained digital ownership &
Informal political influencers are decisive RE stakeholders; trust
must precede design in high-context communities \\
\hline
\textbf{Herluma} &
Urban Canada; designing for anonymity, safety, and trauma in a complex
shelter ecosystem &
Sessions with women with lived homelessness experience; sequential
credibility-building with GVCEH then BC Housing; acknowledgment of
systemic trauma &
Entire design premise invalidated through co-creation: mobile app replaced
by web tool for operators; anonymity and pseudonymization built into
core requirements &
Shelter operators became system advocates and champions of adoption
across the network; privacy controls gave staff ownership &
Co-creation's most important contribution is often the requirements it
eliminates; adoption follows genuine ownership \\
\hline
\textbf{BridgingRoots} &
Remote Arctic Canada; co-designing with a community whose knowledge
systems differ fundamentally from the engineering team's &
18 months of sustained engagement; informal youth interactions before
any design activity; cultural sensitivity to historical extractive
relationships &
Community's oral traditions and land-based knowledge became direct
functional requirements; value conflicts resolved through collective
deliberation; teacher verification embedded in design &
Youth became authors of cultural content; community owns and maintains
a growing body of language knowledge; three youth pursuing higher
education &
Time spent building relationships before requirements are discussed is
not pre-project activity; it is the first RE activity \\
\hline
\end{tabular}
\end{table*}

\subsection{Member Checking Reflection}
\label{sec:validation}

As discussed in Section \ref{sec:memberchecking}, member checking was conducted to examine whether REConnect's three principles reflected community partners' subjective experiences of what enabled successful requirements work and sustained system adoption. Across the member checking responses, community partners confirmed that the principles of building trusting relationships, co-creating with and alongside users, and empowering users as agents of change resonated with their experiences. Importantly, partners did not describe these principles as isolated activities. They described them as mutually reinforcing conditions through which requirements work became meaningful and impactful. 

Firstly, member checking confirmed that community partners experienced these projects as shared initiatives rather than conventional, transactional software handoffs. A partner from the Community Transformation Visualization (CTV) project explained that developing the CTV application was \textit{"not a standard technical handover where a developer  deliver a product and disappear"} but rather, \textit{"it felt more like a shared initiative. The three principles you mentioned reflect how we navigated the challenges of making a digital tool work for a community transformation"}. This distinction is central to REConnect. In community-facing software projects, requirements are not simply elicited from stakeholders and translated into technical artifacts; they must be negotiated through relationships among project teams and community members. The CTV partner further emphasized that the three REConnect principles reflected how the group dealt with the challenges of making a digital platform work for the benefit of an ongoing community transformation effort. 

Furthermore, partners confirmed that trust was not merely helpful to the process, but essential. The community partner for the Herluma project described trust as the base of our framework, explaining that \textit{"the trust needs to come at the bottom"} and that it must be \textit{"the solid foundation"}. Without sufficient trust, she cautioned, co-creation can become superficial, and that participants may \textit{"go along to get along,"} accepting design decisions that do not reflect what they truly need. Thus, trust is not only a condition for access to stakeholders, it is what enables honest disagreement, and corrections during requirements analysis. This reflection strengthens REConnect's first principle: building trust through shared presence. 

Member checking also confirmed that co-creation was most meaningful when users and community members shaped the system from within their own cultural, organizational, and practical contexts. In Bridging Roots, the community partner stated that the Inuvialuktun language application was \textit{"entirely co-created students and elders in the communnity. Elders provided support by using the language, helping us record clips to insert into the app and guiding us with the wording, pronunciation and spelling of words or phrases. This also helped to get our staff and students engaged in using the app with their students at all grade levels, including the Inuvialuktun Language Teacher, who used it with students during her classes"}. This confirms that the requirements of a system emerge through community knowledge, culture, and the participation of those who use and sustain the application, highlighting the necessity of co-creation in RE work.

Community partners also confirmed that empowerment was experienced not simply as increased use of the software, but as increased agency, ownership, and capacity to act. In the Livelihood Tracking Software (LTS), the solution empowered women's self help groups by enabling effective monitoring, evaluation and learning of groups' livelihood activities. The community partner explained, \textit{"Our partners are motivated to bring their livelihood data such as cash flow details, money invested and crop sales from their respective groups as they are able to see their strengths, trends, and motivate the household for financial freedom."} In Bridging Roots, the community partner described \textit{"users and stakeholders were most definitely empowered as \textbf{agents of change} in pushing forward acts of reconciliation and language revitalization}". The application empowered young users by making Inuvialuktun learning both engaging and accessible, while creating a resource that could be used by students, staff, Elders, visitors, and future generations. In the CTV project, empowerment was described through users' \textit{"ability to monitor their networks and assess their growth in nine areas"} of community transformation. In Herluma, empowerment was framed through choice. The community partner explained that those in vulnerable situations often do not realize they have choices, saying \textit{"When you're at a in a very vulnerable State. You don't necessarily understand that you actually do have choice."} This expands REConnect's understanding of empowerment. In vulnerable community contexts, empowerment is not only about ownership of the system, but also about ensuring that the system increases users' ability to understand and evaluate their own contexts.  

Additionally, member checking also provided evidence for the interdependence of all three principles. The CTV partner described this relationship directly: \textit{"I believe, three principles function interdependently because the initial building of trust and rapport among our diverse stakeholders created a safe space that made co-creation possible which allowed coordinators and facilitators from different backgrounds to provide honest feedback that directly shaped a user-friendly of the system. This collaborative design process helped to ensure that the users did not view the application as a complex external imposition but rather as a practical tool they fully understood, which directly helped them to act as empowered agents of change"}. In the Bridging Roots project, co-creation helped users to feel empowered. When asked about its significance, they mentioned that \textit{"co-creating with youth and elders, learning from computer design specialists on how to build an application, creates real relationships and friendships, while empowering everyone to learn and try new things together. It builds on community values of sharing and learning by creating a tangible product that can be used by different generations and to be used in perpetuity by people of all backgrounds."} In LTS project, the community partner connected empowerment directly to sustained engagement: \textit{"We feel empowered due to the team's eagerness and the many hours they spent co-designing with us and training us. As a result, we can train our partners in using the system to continually evaluate and monitor the change in women's self help groups."} Together these reflections support REConnect's central structure: trust creates the relational conditions for co-creation, co-creation produces systems that reflect local values, and empowerment sustains community ownership beyond the project team's involvement.

\section{Discussion}
\label{sec:discussion}

REConnect, a deliberate play on words, re-centers human connection in
requirements engineering activities for social sustainability. Inspired by projects with tangible
societal impact, it emphasizes developing trusting relationships and
sustained engagement with technology end users and stakeholders. This
stands in contrast and in response to recent automation-driven approaches that, while
efficient, risk overlooking the lived experiences and values of the
communities where impact is sought.

A discussion of such extensive emphasis on human
involvement is pertinent and particularly important in an era of rapid advances in AI. The
projects presented in this paper were conducted without AI assistance
in any RE activity; the principles and REActions of REConnect as
described here are grounded entirely in human-centered, participatory
practice. What follows is our reflection on REConnect's distinguishing features relative to related works in participatory design and value-driven RE, as well as how AI might complement,
rather than replace, this approach in emerging approaches to social sustainability projects.

\subsection{REConnect: A Participatory RE approach that matters}

The principles of REConnect treat stakeholders as collaborators, not informants. Drawing on value-based requirements engineering \cite{thew2018value} and community-engaged software engineering \cite{whittle2014citizen}, REConnect grounds solution design in an understanding of the lived experiences and values of the communities where impact is sought. Importantly, REConnect is not intended to replace established requirements engineering best practices, such as conducting systematic stakeholder analyses, carefully documenting requirements, or maintaining traceability throughout development. Rather, it \textit{augments} these practices by foregrounding the relational, value-sensitive dimensions of requirements work. In doing so, REConnect supports a human-centered, and participatory RE approach that ensures systems address both immediate functional needs while also supporting a community's well-being over time. Participation is therefore elevated from episodic feedback to continuous dialogue due to shared authorship, based on trust, engagement of stakeholders in design, and the empowerment of the community.  


REConnect's novel contribution is the introduction of three principles that are interconnected and mutually reinforcing and that, based on our research, form the basis of successful social sustainability projects. At the core of social sustainability is the long-term, sustained outcome of a project in ways that matter to the respective communities. The principle of \emph{empowering end-users and stakeholders as agents of change} is therefore fundamental, and is closely tied to the other two principles of \emph{building trusting relationships} and \emph{co-creating with and alongside end-users}.

\emph{Empowering key stakeholders} is both foundational to, and an outcome of, the REConnect Actions involved in \emph{building trusting relationships}. The process of identifying stakeholders and understanding their roles, motivations and influence within communities helps recognize and validate their lived experiences, expertise, and unique perspectives, fostering a genuine sense of being heard and valued. This led to trust in the authenticity of the engagement process and motivates continued participation in developing solutions that strengthen stakeholders' roles, influence, and sense of ownership within their communities. As one community partner described, the three principles \emph{“function interdependently”} because trust and rapport created the safe conditions that made \emph{co-creation} possible; in turn, this collaborative design process helped users understand the system not as \emph{“a complex external imposition”} but as a practical tool through which they could act as \emph{“empowered agents of change.”} 

As stakeholders contributed meaningfully to the design and implementation of solutions that reflect their values, priorities, and experiences, they became further empowered and increasingly invested in the project's long-term success and sustained outcomes, recognizing these outcomes as a direct result of their participation. This connection was also evident in partners' reflections on capacity building, where one community partner explained that they felt empowered because of the team's \emph{“eagerness”} and the many hours spent \emph{“co-designing with us and training us,”} which enabled them to train their own partners to use the system for continued evaluation and monitoring. Here, co-creation did not only improve the immediate design of the software; it strengthened the community's capacity to sustain, adapt, and govern the system after the project team's involvement.

This emphasis on participation leading to long-term, sustained outcomes is an advancement from traditional Participatory Design  approaches. While early PD approaches treated learning from and empowering participants as important project outcomes, over time PD has given greater attention to the design process itself, with methods often becoming the primary outcome; in contrast, this has resulted in less regard for sustained relationships and outcomes after the project \cite{BodkerKyng2018}. In REConnect, empowerment appears to have contributed directly to long-term success, as illustrated by sustained outcomes in the involved communities. Rather than treating empowerment as a desirable byproduct of participation, REConnect positions empowerment as a requirements engineering concern: a traceable consequence of building trust, co-creating with stakeholders, and designing systems that communities are able and motivated to sustain.

\subsection{The Technology-Value Nexus in Social Sustainability}

The INSPIRE projects for social sustainability make visible how deeply technologies are shaped by the values of the communities they are meant to serve. Across the three illustrative examples, technology is not treated as value-neutral \cite{gabriel2022valuealignment}, but as something that embeds values understood by developers and negotiated with the communities subject to technological impact. In the Herluma project, this is reflected in the protection of the dignity of individuals experiencing homelessness; in BloodSync, in the importance of integrity and credibility of the ward leaders within rural Nepalese communities; and in the Canadian Arctic project, in the commitment to collective and participatory decision-making within Inuit communities. These values are inherently local and contextual, shaped through shared histories, lived experiences, religious beliefs, and cultural traditions.

REConnect comprises the principles and actions that help identify the values that matter within particular communities. These principles serve as guidelines on how to engage meaningfully with stakeholders, not simply as participants, but as the people who shape and interpret those values. In this way, stakeholders are understood as both sources of values and as meaning-makers. Their perspectives, experiences, and cultural contexts help define what is important, ethical, and socially valuable in the development and use of technology. 

In doing so, REConnect advances research in value-based RE \cite{thew2018value}. Value-Based RE provides a rigorous prior treatment of stakeholder values within the RE lifecycle, offering a taxonomy and systematic elicitation methods. It includes, for example, trust as a
stakeholder value to be elicited, but does not treat it as a longitudinal
relational prerequisite that must be cultivated before requirements
discovery becomes possible. More broadly,
VBRE was developed for organizational settings with formally identifiable
stakeholders and does not address informal community influencers, sustained
engagement, or empowerment as an RE outcome. VBRE answers the question
of what values to elicit; REConnect provides guidance on how
to build the relationships critical for those values to surface in system development.

\subsection{Towards AI-Enhanced Participatory RE}


The technology-value nexus should be central to considerations of AI-based systems in our societies. AI-assisted RE tools have demonstrated value in processing large volumes
of stakeholder feedback at scale \cite{groen2017crowd, maalej2025automated},
generating draft requirements artifacts \cite{ronanki2023chatgpt,
almeida2024generative}, and surfacing latent patterns in unstructured
data \cite{ghosh2024exploring}. These capabilities address a genuine
bottleneck in participatory RE: the volume of qualitative data generated
through immersive field work, co-design workshops, and community
consultations can exceed what a small research team can analyze manually.
However, AI tools trained on existing data inherit the biases and
exclusions of that data \cite{ferrara2023fairness, tronnier2024bias},
thus their outputs require human interpretation to be contextually
meaningful \cite{mitchell2025fully}. As LLMs move
toward agentic configurations capable of autonomous requirements
elicitation \cite{mukherjee2025agentic} and generating design options, the risk of systematically
filtering out community values, informal influences, and historically
marginalized voices intensifies.


REConnect provides a participatory approach that brings forth the relational conditions under which AI can support software practioners' requirements work without displacing the human judgement of those impacted by the solution, or the trust and community ownership on which socially sustainable software depends. We discuss how software practitioners can engage with stakeholders, including end-users and community partners, to maintain their human agency during software design by contributing values and requirements, and by actively evaluating and refining AI-generated outputs. This human stakeholder-centered approach leverages  emerging roles such as Forward Deployed Engineers \cite{kim2026forward}, who are software engineers increasingly positioned within AI development organizations to conduct "product" and "feature discovery" in close collaboration with users. REConnect highlights how such roles can act genuinely within participatory, stakeholder-centred requirements practices that preserve human agency while leveraging AI support.

The following section  describes each step of REConnect and how its stakeholder - software practitioner collaboration is supported by AI automation. By software practitioner we refer to any member of the software development team using AI-assistants for requirements or design artifact generation, such as FDEs or designers or developers. 
Table \ref{tab:stakeholder_roles} summarizes stakeholders' roles in human–AI requirements-centered collaboration and the strategies used to engage and empower them at each stage.

\begin{table*}[h]
\centering

\caption{Stakeholder Roles in Human--AI requirements-centered Collaboration  and Engagement leading to Empowerment Strategies}\label{tab:stakeholder_roles}
\begin{tabular}{p{3cm} p{3cm} p{4.5cm} p{5cm}}

\toprule
\textbf{REConnect } & \textbf{Stakeholder Role} & \textbf{Description in Human--AI Collaboration} & \textbf{Engagement \& Empowerment Strategies} \\ \midrule

\textbf{Step 1. Human-Centered Groundwork} & Guardians of values and ethics & Set ethical guardrails (e.g., fairness, accessibility) and monitor alignment. & \textbf{Value alignment} - Ensure requirements reflect stakeholders’ moral and social priorities. \\
\midrule
\textbf{Step 2. AI-Assisted Knowledge Discovery} & Curators and contextualisers of AI outputs & Define AI goals, boundaries, and evaluation criteria. Interpret AI findings in light of domain expertise and lived experience. & \textbf{Influence} - Shape which insights are considered relevant or actionable. \\
\midrule
\multirow{2}{*}{\parbox{3cm}{\textbf{Step 3. Human Interpretation \& Sensemaking}}} 
& Curators and contextualisers of AI outputs & Interpret AI-generated outputs in workshops, challenge or refine insights. & \textbf{Influence} - Contextualise findings with lived experience. \\
& Inclusion amplifiers & Identify missing perspectives and guide targeted outreach. & \textbf{Representation} - Ensure diverse voices are integrated. \\
\midrule
\textbf{Step 4. Aligning Stakeholder Motivations to Capabilities} & Accountability partners & Monitor traceability from stakeholder values to design elements. & \textbf{Transparency \& trust} - Verify how input is used, reinforcing trust in the process. \\
\midrule
\multirow{2}{*}{\parbox{3cm}{\textbf{Step 5. Co-Creating with AI: Iterative Prototyping}}} 
& Co-reflectors in iterative review cycles & Engage in reflective workshops to refine AI-generated prototypes. & \textbf{Autonomy} - Influence over the design of the process itself, not just system output. \\
&   &   & \textbf{Ownership} - Shape requirements and designs over multiple iterations. \\
\bottomrule
\end{tabular}
\end{table*}

\textbf{Step 1. Human-Centered Groundwork}. By enacting the REActions of Building Relationships and Trust with Stakeholders, practitioners first identify key stakeholder roles, community influencers, power dynamics, and cultural or political factors that shape the project environment and stakeholders' values and motivations. They may create rich stakeholder maps or personas in collaboration with domain experts. An organization should consider not only formal roles, such as a hospital administrator, but also informal influencers, such as a community volunteer, a trusted leader, or a respected elder. Personas and contextual understanding are critical because, depending on the software project, the influencers could be decisively different. Crucially, this stage involves in-person presence and participation in the stakeholders’ environment, observing daily routines, and engaging in open-ended conversations. Through these interactions, requirement practitioners can glean subtle contextual details. For example, important details could revolve around community norms and personal trauma related to community events, which would unlikely surface from automated or online-based analysis alone. 

Trust and cultural understanding are inherently social and experiential. Based on current technology, AI cannot replace the experience of a requirements practitioner drinking tea around a village fire to hear concerns, or a team shadowing homeless shelter staff through a busy afternoon shift to feel their stress. In this step, humans articulate realities of their lived experiences, and become \emph{meaning-makers}.
They decide what matters in their community or organization. The result of this step is a richer understanding of 'who to listen to' (formal and informal stakeholder roles and influencers) and 'why' (their values, motivations) in the unique context of the project. Similar to how guardrails provide DevOps engineers with predefined policies and checks to improve safety and security \cite{kim2021devops}, the insights from this step allow the humans to \emph{define ethical guardrails} (e.g., fairness, cultural alignment, inclusiveness). In an elevated role as \emph{guardians of values and ethics}, stakeholders can later use these guardrails during requirements definition or co-design activities by ensuring requirements and design reflect stakeholders' moral and social priorities. Furthermore, the relationships developed through this initial groundwork lead to key stakeholder buy-in and engagement in later co-design stages in REConnect. \\

\textbf{Step 2. AI-Assisted Knowledge Discovery of Stakeholder Input}. After developing an understanding of the stakeholder lived and organizational ecosystem, the REConnect process can leverage generative AI to improve the efficiency of the discovery of requirements-relevant knowledge. In practice, this step might involve aggregating and analyzing large volumes of unstructured stakeholder input that could be overwhelming for manual analysis, for example, thousands of user reviews, support tickets, social media discussions, or transcripts from stakeholder workshops. LLM models can be used to sift through this data and identify patterns, emerging concerns, and latent needs that stakeholders have expressed. For example, a generative model can summarize user feedback or automatically highlight frequent grievances. 

Moreover, this AI-assisted knowledge discovery is not limited to online feedback insights. It can be leveraged to handle stakeholder input from other sources such as interviews, workshops, and surveys. After automated transcription of stakeholder input, organizations can use generative models to generate insights from the data. This AI-assisted knowledge discovery aligns with CrowdRE's goals of listening to the crowd at scale, but with more semantic consideration. The RE team can prompt an LLM to focus on distinct concerns from various user groups. 

It is important to highlight that the AI’s role in this step is supportive and exploratory. The goal is not to have the AI make final decisions about requirements, but to enhance RE practitioner awareness. By automating the synthesis of large datasets, requirements practitioners can focus more on the interpretation of the collected data.

\textbf{Step 3. Human Interpretation and Sensemaking in Context.} 
After AI-assisted knowledge discovery, the next stage involves 
interpreting AI outputs through the contextual knowledge and lived 
experience gained in Stage 1. In collaborative workshops, stakeholders 
and requirements practitioners review emerging needs and insights, 
challenging, refining, or contextualizing AI-generated artifacts. 
Humans retain interpretive authority: they can correct 
misinterpretations, fill contextual gaps, and ensure AI-generated 
summaries reflect the community's voices and actual priorities rather than the 
patterns most visible in the data.
Stakeholders act as \emph{inclusion amplifiers}, identifying missing 
perspectives and guiding targeted follow-up mining or stakeholder interaction. This is perhaps one of the most important reasons and mechanisms for maintaining human agency for an inclusive requirements discovery in an AI-assisted software development process. Online sources are not a neutral or complete source of requirements; they are subject to sampling bias. For example, Blincoe and Tizard's \cite{TizardRLB22} report statistically significant demographic differences in who contributes online user reviews, and some groups (women) were underrepresented in the data. In such scenarios, stakeholders, drawing on contextual and domain knowledge, can highlight this gap and prompt a broader elicitation, through 1) further prompting of LLMs to surface not only dominant, highly visible themes but also less obvious or underrepresented concerns (in this case from women), or 2) further elicitation from stakeholders including women through interviews or collaborative workshops. 

Similarly, an AI summarization may fail to capture a community's historical 
mistrust of certain institutions, or that a trusted community leader 
matters more for delivering notifications than an impersonal app; 
stakeholders can flag these omissions as critical, surfacing design 
considerations that would remain invisible without their contextual 
knowledge. This human-centric synthesis step acts as quality control 
between AI interpretation and requirements formulation: AI contributes 
input, but humans retain their role as decision-makers rather than 
passive recipients.

\textbf{Step 4. Aligning Stakeholder Motivations to Solution Capabilities}. A critical aspect of REConnect is the emphasis on empowering stakeholders as agents of sustained change in their communities, which entails aligning the software’s design with the intrinsic motivations of the people who will use it. During initial co-designing activities, requirements practitioners and stakeholders collaboratively map stakeholder motivations to envisioned system capabilities. This is a creative activity drawing on the trust and understanding developed earlier. Each influential stakeholder or group (identified back in Step 1) is considered. Key questions that a requirements practitioner should answer include: What drives the stakeholder? How could the system enable them, address their concerns, or unlock new possibilities to solve their problems? For example, in the BloodSync case, we recognized that ward leaders were motivated by community duty and a desire to prove themselves as champions of public welfare. Correspondingly, a software feature was designed to allow ward leaders to coordinate local blood donors. This feature harnessed their motivation for leadership into a feature of the system. This mapping process requires empathy and often a negotiation of value trade-offs. 

In this step, AI plays a little direct role because, as of this writing, it cannot determine stakeholders' motivations rooted in cultural, political, or socio-economic contexts. The output of this step is a list of stakeholders and how they might contribute to solution co-creation activities because of their motivation to changes in the problem domain; most importantly are those who are identified as informal influencers in their organization or community. People are more likely to champion a system if they see their intentions and values integrated into its software capabilities.

\textbf{Step 5. Co-creation with AI: Iterative Prototyping}. With a validated set of requirements and understanding of stakeholder values and motivations, REConnect engages the humans in co-creation activities, which can be enhanced by AI-generated prototypes. This step extends our earlier REActions of ``Iterative Prototyping and Simulation" to rapidly produce design artifacts, and involve stakeholders as \emph{co-reflectors in iterative review and refinement of solution designs}. Using the requirements and context as input, a generative AI tool could generate initial prototypes of the solution. For a user interface, this might involve creating wireframe ideas or even visual mock-ups by leveraging generative visual models or UI-specific generation tools. Additionally, a generative AI model may draft workflow diagrams or storyboards of user interactions. For textual artifacts, an LLM could generate sample user stories, use case descriptions, or even pseudo-code demonstrating how a requirement might be implemented.

One advantage of generative AI in this step is speed. Since new prototypes or variants can be generated in minutes, the practitioners can iterate through feedback loops much faster. In a workshop, a group of stakeholders might reject an AI-generated interface as too complex; the facilitator can immediately adjust the prompt for the relevant generative AI model and produce a revised version on the spot for the group to consider. Generative AI tools currently allow stakeholders to directly shape the evolving design without needing specialized skills in drawing or coding \cite{Liu2025HumanAICoCreation}.

Taken together, the five stages outline how AI can be embedded within a relational approach to RE within the REConnect framework. 
AI contributes efficiency at the knowledge discovery and prototyping 
stages, but the stages that determine whether software serves social 
sustainability, understanding who holds power in a community, surfacing 
historical mistrust, mapping motivations to capabilities, remain 
irreducibly human. In the INSPIRE corpus, the requirements that 
mattered most, that a ward leader's buy-in was a prerequisite for 
adoption, that anonymity was non-negotiable for shelter users, that 
oral traditions were functional requirements, could not have been 
produced by any automated process. REConnect's principles, 
building trust, co-designing with communities, and empowering users, are not constraints on AI integration but are criteria against which any AI-assisted RE 
activity must be evaluated if the goal is software that communities 
adopt, sustain, and own.

\subsection{Threats to Validity}

\textbf{Construct validity.} The three REConnect principles were derived through reflective synthesis followed by thematic analysis, conducted by researchers who were also
participants in the processes they studied, creating a risk that the
constructs reflect the researchers' interpretive lens rather than the
phenomena themselves. We acknowledge, however, that in reflective synthesis as a research method, the researchers' engagement with the experience is treated as a source of knowledge, rather than as a threat to validity \cite{fook2011developing}. We mitigated possible biases in our thematic analysis by independently coding prior to joint discussion, grounding interpretations in publicly documented project artifacts wherever possible, and member checking with community partner representatives, which provides a credibility check on whether the principles resonate with communities' own accounts of what mattered.

\textbf{Internal validity.} The paper claims that trust, co-design, and
empowerment are the conditions that enabled successful social sustainability
outcomes across the INSPIRE corpus. However, the causal links between
these principles and outcomes such as adoption and community well-being
are implied rather than formally established. The evidence base is
observational and interpretive rather than experimental, and the
consistent pattern across 26 projects supports the plausibility of the
relationship but does not establish it causally. Future work employing
controlled comparison or longitudinal outcome measurement would strengthen
this claim.

\textbf{External validity.} REConnect was derived within a specific
program context: community-engaged software projects conducted through
an academic experiential learning program with student teams, addressing
social sustainability challenges in two regional settings. The framework
is most applicable to socio-techical, community-facing software projects
where sustained engagement is feasible and social sustainability outcomes
are sought. It is not claimed as a universal RE framework: contexts with
tight delivery schedules, enterprise stakeholder structures, or large
and anonymous user bases may require different or complementary
approaches. The 26-project corpus provides variation across community
type, geographic setting, and social sustainability domain, supporting
analytic generalizability within the class of community-facing software
projects, but not beyond it.

\textbf{Reliability.} The primary data sources, including 
project artifacts, student and team retrospectives, interview transcripts, and
debrief records, were generated by student project teams, community
partners, and mentors across the four-year program. This introduces
variability in documentation depth and consistency across projects,
mitigated by triangulating across multiple artifact types per project
and by the authors' sustained involvement as program coordinators. The
public availability of project retrospectives on the INSPIRE program
website provides an auditable record that partially offsets this
limitation.

\section{Conclusion and Future Work}

Software that contributes to social sustainability cannot be built at
a distance from the communities it serves. This paper has argued that
requirements engineering carries a particular responsibility in
community-facing software development that automation-driven approaches
cannot fulfill alone. CrowdRE and AI-assisted elicitation scale well
but systematically exclude the informal influencers, cultural dynamics,
and trust-dependent knowledge that determine whether a system will be
adopted and sustained by the community it serves.

REConnect responds to this gap by re-centering RE on human connection.
Derived through inductive thematic analysis across 26 community-engaged
software projects, and illustrated through three projects spanning rural Nepal,
urban British Columbia, and remote Arctic Canada, REConnect organizes
its contribution around three principles: building trusting relationships
as a prerequisite for substantive requirements discovery; co-creating
with and alongside users as the mechanism through which formally
specifiable, culturally grounded requirements are produced; and
empowering users as agents of change as the condition through which
systems achieve durable community ownership. Each principle is
operationalized through a set of REActions that embed these commitments
across the RE lifecycle from elicitation through post-delivery engagement.

Member checking with community partner representatives provides initial
evidence that the three principles resonate with communities' own
accounts of what enabled successful requirements work and system
adoption. A comparative analysis against participatory design,
value-based RE, and CBPR indicates that no prior framework
or method addresses all three principles in combination within a formal
RE lifecycle for low-resource and marginalized community contexts,
though we acknowledge that this comparative framing is our own and
warrants scrutiny through independent replication and peer evaluation.\\

\textbf{\textit{Future Work.}} This paper intends to bring back, and keep the human element back in the requirements engineering activities of software development. As AI continues to dominate SE, human agency will undoubtedly become one of the most important focal points for our field. Thus, the growing trends towards roles such as Forward Deployed Engineers and End-user Software Engineering \cite{robinson2025requirements} imply that software development becomes more accessible to non-technical roles, providing requirements engineering research with both unprecedented opportunities and responsibilities. Focusing research efforts on advancing our understanding, methods and approaches to aligning software with human and societal values will be worthwhile to prevent AI-mediated RE from reproducing the extractive relationships that lead to software that is unfair, discriminatory or not inclusive. The causal relationship between REConnect practices and social sustainability outcomes has not been formally
established and calls for longitudinal measurement beyond the observational evidence presented here. We hope that REConnect provides a useful framework to guide future studies that bring further validation and advancement on these principles, especially within the AI integrated agenda we outlined in our paper. 





\bibliographystyle{cas-model2-names}

\bibliography{cas-refs}

\begin{thebibliography}{100}
\expandafter\ifx\csname natexlab\endcsname\relax\def\natexlab#1{#1}\fi
\providecommand{\url}[1]{\texttt{#1}}
\providecommand{\href}[2]{#2}
\providecommand{\path}[1]{#1}
\providecommand{\DOIprefix}{doi:}
\providecommand{\ArXivprefix}{arXiv:}
\providecommand{\URLprefix}{URL: }
\providecommand{\Pubmedprefix}{pmid:}
\providecommand{\doi}[1]{\href{http://dx.doi.org/#1}{\path{#1}}}
\providecommand{\Pubmed}[1]{\href{pmid:#1}{\path{#1}}}
\providecommand{\bibinfo}[2]{#2}
\ifx\xfnm\relax \def\xfnm[#1]{\unskip,\space#1}\fi
\bibitem[{Aizenberg and van~den Hoven(2020)}]{aizenberg2020designing}
\bibinfo{author}{Aizenberg, E.}, \bibinfo{author}{van~den Hoven, J.}, \bibinfo{year}{2020}.
\newblock \bibinfo{title}{Designing for human rights: Bridging the gap between values and design requirements}.
\newblock \bibinfo{journal}{arXiv preprint arXiv:2005.04949} \URLprefix \url{https://arxiv.org/abs/2005.04949}.
\bibitem[{Ajmal et~al.(2018)Ajmal, Khan, Hussain and Helo}]{ajmal2018conceptualizing}
\bibinfo{author}{Ajmal, M.M.}, \bibinfo{author}{Khan, M.}, \bibinfo{author}{Hussain, M.}, \bibinfo{author}{Helo, P.}, \bibinfo{year}{2018}.
\newblock \bibinfo{title}{Conceptualizing and incorporating social sustainability in the business world}.
\newblock \bibinfo{journal}{International Journal of Sustainable Development \& World Ecology} \bibinfo{volume}{25}, \bibinfo{pages}{327--339}.
\bibitem[{Al-kfairy and et~al.(2024)}]{alkfairy2024ethical}
\bibinfo{author}{Al-kfairy, H.}, \bibinfo{author}{et~al.}, \bibinfo{year}{2024}.
\newblock \bibinfo{title}{Ethical considerations of generative ai: Systematic review}.
\newblock \bibinfo{journal}{Informatics} \bibinfo{volume}{11}, \bibinfo{pages}{58}.
\newblock \URLprefix \url{https://www.mdpi.com/2227-9709/11/3/58}.
\bibitem[{Alidoosti et~al.(2023)Alidoosti, De~Sanctis, Iovino, Lago and Razavian}]{alidoosti2023stakeholder}
\bibinfo{author}{Alidoosti, R.}, \bibinfo{author}{De~Sanctis, M.}, \bibinfo{author}{Iovino, L.}, \bibinfo{author}{Lago, P.}, \bibinfo{author}{Razavian, M.}, \bibinfo{year}{2023}.
\newblock \bibinfo{title}{Stakeholder inclusion and value diversity: An evaluation using an access control system}, in: \bibinfo{booktitle}{European Conference on Software Architecture}, \bibinfo{organization}{Springer}. pp. \bibinfo{pages}{71--88}.
\bibitem[{Almeida and Silva(2024)}]{almeida2024generative}
\bibinfo{author}{Almeida, M.}, \bibinfo{author}{Silva, M.}, \bibinfo{year}{2024}.
\newblock \bibinfo{title}{Generative ai for requirements engineering: A systematic literature review}.
\newblock \bibinfo{journal}{arXiv preprint arXiv:2409.06741} \URLprefix \url{https://arxiv.org/abs/2409.06741}.
\bibitem[{Ataei et~al.(2024)Ataei, Cheong, Grandi, Wang, Morris and Tessier}]{ataei2024elicitron}
\bibinfo{author}{Ataei, M.}, \bibinfo{author}{Cheong, H.}, \bibinfo{author}{Grandi, D.}, \bibinfo{author}{Wang, Y.}, \bibinfo{author}{Morris, N.}, \bibinfo{author}{Tessier, A.}, \bibinfo{year}{2024}.
\newblock \bibinfo{title}{Elicitron: An llm agent-based simulation framework for design requirements elicitation}.
\newblock \bibinfo{journal}{arXiv preprint arXiv:2404.16045} \URLprefix \url{https://arxiv.org/abs/2404.16045}.
\bibitem[{Autili et~al.(2025)Autili, De~Sanctis, Inverardi and Pelliccione}]{autili2025engineering}
\bibinfo{author}{Autili, M.}, \bibinfo{author}{De~Sanctis, M.}, \bibinfo{author}{Inverardi, P.}, \bibinfo{author}{Pelliccione, P.}, \bibinfo{year}{2025}.
\newblock \bibinfo{title}{Engineering digital systems for humanity: A research roadmap}.
\newblock \bibinfo{journal}{ACM Transactions on Software Engineering and Methodology} \bibinfo{volume}{34}, \bibinfo{pages}{1--33}.
\bibitem[{Bano and et~al.(2023)}]{bano2023operationalizing}
\bibinfo{author}{Bano, M.}, \bibinfo{author}{et~al.}, \bibinfo{year}{2023}.
\newblock \bibinfo{title}{Operationalizing diversity and inclusion in ai systems through requirements engineering}.
\newblock \bibinfo{journal}{arXiv preprint arXiv:2311.14695} \URLprefix \url{https://arxiv.org/abs/2311.14695}.
\bibitem[{Bennaceur et~al.(2024)Bennaceur, Ghezzi, Kramer, Nuseibeh, Werthner and Nida-R{\"u}melin}]{bennaceur2024responsible}
\bibinfo{author}{Bennaceur, A.}, \bibinfo{author}{Ghezzi, C.}, \bibinfo{author}{Kramer, J.}, \bibinfo{author}{Nuseibeh, B.}, \bibinfo{author}{Werthner, H.}, \bibinfo{author}{Nida-R{\"u}melin, J.}, \bibinfo{year}{2024}.
\newblock \bibinfo{title}{Responsible software engineering: Requirements and goals}.
\newblock \bibinfo{journal}{Hannes Werthner{\textperiodcentered} Carlo Ghezzi{\textperiodcentered} Jeff Kramer{\textperiodcentered} Julian Nida-R{\"u}melin{\textperiodcentered} Bashar Nuseibeh{\textperiodcentered} Erich Prem{\textperiodcentered}} , \bibinfo{pages}{299}.
\bibitem[{Beyer and Holtzblatt(1998)}]{beyer1998contextual}
\bibinfo{author}{Beyer, H.}, \bibinfo{author}{Holtzblatt, K.}, \bibinfo{year}{1998}.
\newblock \bibinfo{title}{Contextual Design: Defining Customer-Centered Systems}.
\newblock \bibinfo{publisher}{Morgan Kaufmann}, \bibinfo{address}{San Francisco, CA}.
\bibitem[{Beyer and Holtzblatt(1999)}]{beyer1999contextual}
\bibinfo{author}{Beyer, H.}, \bibinfo{author}{Holtzblatt, K.}, \bibinfo{year}{1999}.
\newblock \bibinfo{title}{Contextual design}.
\newblock \bibinfo{journal}{interactions} \bibinfo{volume}{6}, \bibinfo{pages}{32--42}.
\bibitem[{B{\o}dker and Kyng(2018)}]{BodkerKyng2018}
\bibinfo{author}{B{\o}dker, S.}, \bibinfo{author}{Kyng, M.}, \bibinfo{year}{2018}.
\newblock \bibinfo{title}{Participatory design that matters---facing the big issues}.
\newblock \bibinfo{journal}{ACM Transactions on Computer-Human Interaction} \bibinfo{volume}{25}, \bibinfo{pages}{4:1--4:31}.
\newblock \DOIprefix\doi{10.1145/3152421}.
\bibitem[{Boud et~al.(2013)Boud, Keogh and Walker}]{boud2013reflection}
\bibinfo{author}{Boud, D.}, \bibinfo{author}{Keogh, R.}, \bibinfo{author}{Walker, D.}, \bibinfo{year}{2013}.
\newblock \bibinfo{title}{Reflection: Turning experience into learning}.
\newblock \bibinfo{publisher}{Routledge}.
\bibitem[{Braun and Clarke(2006)}]{braun2006thematic}
\bibinfo{author}{Braun, V.}, \bibinfo{author}{Clarke, V.}, \bibinfo{year}{2006}.
\newblock \bibinfo{title}{Using thematic analysis in psychology}.
\newblock \bibinfo{journal}{Qualitative Research in Psychology} \bibinfo{volume}{3}, \bibinfo{pages}{77--101}.
\newblock \DOIprefix\doi{10.1191/1478088706qp063oa}.
\bibitem[{Burnay and Snoeck(2017)}]{burnay2017trust}
\bibinfo{author}{Burnay, C.}, \bibinfo{author}{Snoeck, M.}, \bibinfo{year}{2017}.
\newblock \bibinfo{title}{Trust in requirements elicitation: how does it build, and why does it matter to requirements engineers?}, in: \bibinfo{booktitle}{Proceedings of the Symposium on Applied Computing}, pp. \bibinfo{pages}{1094--1100}.
\bibitem[{Burnett et~al.(2021)Burnett, Fallatah, Hu, Perdriau, Mendez, Gao and Sarma}]{burnett2021toward}
\bibinfo{author}{Burnett, M.}, \bibinfo{author}{Fallatah, A.}, \bibinfo{author}{Hu, C.}, \bibinfo{author}{Perdriau, C.}, \bibinfo{author}{Mendez, C.}, \bibinfo{author}{Gao, C.}, \bibinfo{author}{Sarma, A.}, \bibinfo{year}{2021}.
\newblock \bibinfo{title}{Toward an actionable socioeconomic-aware hci}.
\newblock \bibinfo{journal}{arXiv e-prints} , \bibinfo{pages}{arXiv--2108}.
\bibitem[{Burrows et~al.(2019)Burrows, Lopez-Lorca, Sterling, Miller, Mendoza and Pedell}]{burrows2019motivational}
\bibinfo{author}{Burrows, R.}, \bibinfo{author}{Lopez-Lorca, A.}, \bibinfo{author}{Sterling, L.}, \bibinfo{author}{Miller, T.}, \bibinfo{author}{Mendoza, A.}, \bibinfo{author}{Pedell, S.}, \bibinfo{year}{2019}.
\newblock \bibinfo{title}{Motivational modelling in software for homelessness: Lessons from an industrial study}, in: \bibinfo{booktitle}{2019 IEEE 27th International Requirements Engineering Conference (RE)}, \bibinfo{organization}{IEEE}. pp. \bibinfo{pages}{297--307}.
\bibitem[{Collins et~al.(2018)Collins, Clifasefi, Stanton, Straits, Gil-Kashiwabara, Rodriguez~Espinosa, Nicasio, Andrasik, Hawes, Miller, Nelson, Valez, Holliday, Austin, Watson, Clifford, Doyle and Wallerstein}]{collins2018cbpr}
\bibinfo{author}{Collins, S.E.}, \bibinfo{author}{Clifasefi, S.L.}, \bibinfo{author}{Stanton, J.}, \bibinfo{author}{Straits, K.J.E.}, \bibinfo{author}{Gil-Kashiwabara, E.}, \bibinfo{author}{Rodriguez~Espinosa, P.}, \bibinfo{author}{Nicasio, A.V.}, \bibinfo{author}{Andrasik, M.P.}, \bibinfo{author}{Hawes, S.M.}, \bibinfo{author}{Miller, K.B.}, \bibinfo{author}{Nelson, L.A.}, \bibinfo{author}{Valez, C.N.}, \bibinfo{author}{Holliday, C.N.}, \bibinfo{author}{Austin, L.}, \bibinfo{author}{Watson, E.}, \bibinfo{author}{Clifford, J.}, \bibinfo{author}{Doyle, S.R.}, \bibinfo{author}{Wallerstein, N.}, \bibinfo{year}{2018}.
\newblock \bibinfo{title}{Community-based participatory research ({CBPR}): Towards equitable involvement of community in psychology research}.
\newblock \bibinfo{journal}{American Psychologist} \bibinfo{volume}{73}, \bibinfo{pages}{884--898}.
\newblock \DOIprefix\doi{10.1037/amp0000167}.
\bibitem[{Damodaran(1996)}]{damodaran1996user}
\bibinfo{author}{Damodaran, L.}, \bibinfo{year}{1996}.
\newblock \bibinfo{title}{User involvement in the systems design process-a practical guide for users}.
\newblock \bibinfo{journal}{Behaviour \& information technology} \bibinfo{volume}{15}, \bibinfo{pages}{363--377}.
\bibitem[{Deci and Ryan(2013)}]{deci2013intrinsic}
\bibinfo{author}{Deci, E.L.}, \bibinfo{author}{Ryan, R.M.}, \bibinfo{year}{2013}.
\newblock \bibinfo{title}{Intrinsic motivation and self-determination in human behavior}.
\newblock \bibinfo{publisher}{Springer Science \& Business Media}.
\bibitem[{Dodge and Kitchin(2011)}]{dodge2011code}
\bibinfo{author}{Dodge, M.}, \bibinfo{author}{Kitchin, R.}, \bibinfo{year}{2011}.
\newblock \bibinfo{title}{Code/space: software and everyday life}.
\newblock \bibinfo{journal}{Massachusetts: MIT Press. Dubbeld, L.(2005). The role of technology in shaping CCTV surveillance practices. Information} .
\bibitem[{Duboc et~al.(2019)Duboc, McCord, Becker and Ahmed}]{duboc2019critical}
\bibinfo{author}{Duboc, L.}, \bibinfo{author}{McCord, C.}, \bibinfo{author}{Becker, C.}, \bibinfo{author}{Ahmed, S.I.}, \bibinfo{year}{2019}.
\newblock \bibinfo{title}{Critical requirements engineering in practice}.
\newblock \bibinfo{journal}{IEEE Software} \bibinfo{volume}{37}, \bibinfo{pages}{17--24}.
\bibitem[{Ferrara(2023)}]{ferrara2023fairness}
\bibinfo{author}{Ferrara, E.}, \bibinfo{year}{2023}.
\newblock \bibinfo{title}{Fairness and bias in artificial intelligence: Sources, implications, and mitigation}.
\newblock \bibinfo{journal}{arXiv preprint arXiv:2304.07683} \URLprefix \url{https://arxiv.org/abs/2304.07683}.
\bibitem[{Ferrario et~al.(2016)Ferrario, Simm, Forshaw, Gradinar, Smith and Smith}]{ferrario2016values}
\bibinfo{author}{Ferrario, M.A.}, \bibinfo{author}{Simm, W.}, \bibinfo{author}{Forshaw, S.}, \bibinfo{author}{Gradinar, A.}, \bibinfo{author}{Smith, M.T.}, \bibinfo{author}{Smith, I.}, \bibinfo{year}{2016}.
\newblock \bibinfo{title}{Values-first se: research principles in practice}, in: \bibinfo{booktitle}{Proceedings of the 38th international conference on software engineering companion}, pp. \bibinfo{pages}{553--562}.
\bibitem[{Ferrario et~al.(2014)Ferrario, Simm, Newman, Forshaw and Whittle}]{ferrario2014speedplay}
\bibinfo{author}{Ferrario, M.A.}, \bibinfo{author}{Simm, W.}, \bibinfo{author}{Newman, P.}, \bibinfo{author}{Forshaw, S.}, \bibinfo{author}{Whittle, J.}, \bibinfo{year}{2014}.
\newblock \bibinfo{title}{Software engineering for `{Social Good}': Integrating action research, participatory design, and agile development}, in: \bibinfo{booktitle}{Companion Proceedings of the 36th International Conference on Software Engineering}, pp. \bibinfo{pages}{520--523}.
\newblock \DOIprefix\doi{10.1145/2591062.2591121}.
\bibitem[{Floridi(2015)}]{floridi2015onlife}
\bibinfo{author}{Floridi, L.}, \bibinfo{year}{2015}.
\newblock \bibinfo{title}{The onlife manifesto: Being human in a hyperconnected era}.
\newblock \bibinfo{publisher}{Springer nature}.
\bibitem[{Fook(2011)}]{fook2011developing}
\bibinfo{author}{Fook, J.}, \bibinfo{year}{2011}.
\newblock \bibinfo{title}{Developing critical reflection as a research method}, in: \bibinfo{booktitle}{Creative spaces for qualitative researching: Living research}. \bibinfo{publisher}{SensePublishers Rotterdam}, pp. \bibinfo{pages}{55--64}.
\bibitem[{Friedman et~al.(2013)Friedman, Kahn~Jr, Borning and Huldtgren}]{friedman2013value}
\bibinfo{author}{Friedman, B.}, \bibinfo{author}{Kahn~Jr, P.H.}, \bibinfo{author}{Borning, A.}, \bibinfo{author}{Huldtgren, A.}, \bibinfo{year}{2013}.
\newblock \bibinfo{title}{Value sensitive design and information systems}, in: \bibinfo{booktitle}{Early engagement and new technologies: Opening up the laboratory}. \bibinfo{publisher}{Springer}, pp. \bibinfo{pages}{55--95}.
\bibitem[{Friedman and Nissenbaum(1996)}]{friedman1996bias}
\bibinfo{author}{Friedman, B.}, \bibinfo{author}{Nissenbaum, H.}, \bibinfo{year}{1996}.
\newblock \bibinfo{title}{Bias in computer systems}.
\newblock \bibinfo{journal}{ACM Transactions on information systems (TOIS)} \bibinfo{volume}{14}, \bibinfo{pages}{330--347}.
\bibitem[{Gabriel and Ghazavi(2022)}]{gabriel2022valuealignment}
\bibinfo{author}{Gabriel, I.}, \bibinfo{author}{Ghazavi, V.}, \bibinfo{year}{2022}.
\newblock \bibinfo{title}{The challenge of value alignment: From fairer algorithms to ai safety}, in: \bibinfo{editor}{V{\'e}liz, C.} (Ed.), \bibinfo{booktitle}{The Oxford Handbook of Digital Ethics}. \bibinfo{publisher}{Oxford University Press}, pp. \bibinfo{pages}{336--355}.
\newblock \URLprefix \url{https://doi.org/10.1093/oxfordhb/9780198857815.013.18}, \DOIprefix\doi{10.1093/oxfordhb/9780198857815.013.18}.
\bibitem[{Gama(2024)}]{gama2024awareness}
\bibinfo{author}{Gama, K.}, \bibinfo{year}{2024}.
\newblock \bibinfo{title}{On the awareness about diversity and inclusion being integrated to requirements engineering}, in: \bibinfo{booktitle}{2024 IEEE/ACM Workshop on Multi-disciplinary, Open, and RElevant Requirements Engineering (MO2RE)}.
\bibitem[{Ghosh et~al.(2024)Ghosh, Pargaonkar and Eisty}]{ghosh2024exploring}
\bibinfo{author}{Ghosh, T.K.}, \bibinfo{author}{Pargaonkar, A.}, \bibinfo{author}{Eisty, N.U.}, \bibinfo{year}{2024}.
\newblock \bibinfo{title}{Exploring requirements elicitation from app store user reviews using large language models}.
\newblock \bibinfo{journal}{arXiv preprint arXiv:2409.15473} \URLprefix \url{https://arxiv.org/abs/2409.15473}.
\bibitem[{Graf-Vlachy et~al.(2018)Graf-Vlachy, Buhtz and K{\"o}nig}]{graf2018social}
\bibinfo{author}{Graf-Vlachy, L.}, \bibinfo{author}{Buhtz, K.}, \bibinfo{author}{K{\"o}nig, A.}, \bibinfo{year}{2018}.
\newblock \bibinfo{title}{Social influence in technology adoption: taking stock and moving forward}.
\newblock \bibinfo{journal}{Management Review Quarterly} \bibinfo{volume}{68}, \bibinfo{pages}{37--76}.
\bibitem[{Groen et~al.(2017)Groen, Seyff, Ali, Dalpiaz, Doerr, Guzman, Hosseini, Marco, Oriol, Perini et~al.}]{groen2017crowd}
\bibinfo{author}{Groen, E.C.}, \bibinfo{author}{Seyff, N.}, \bibinfo{author}{Ali, R.}, \bibinfo{author}{Dalpiaz, F.}, \bibinfo{author}{Doerr, J.}, \bibinfo{author}{Guzman, E.}, \bibinfo{author}{Hosseini, M.}, \bibinfo{author}{Marco, J.}, \bibinfo{author}{Oriol, M.}, \bibinfo{author}{Perini, A.}, et~al., \bibinfo{year}{2017}.
\newblock \bibinfo{title}{The crowd in requirements engineering: The landscape and challenges}.
\newblock \bibinfo{journal}{IEEE software} \bibinfo{volume}{34}, \bibinfo{pages}{44--52}.
\bibitem[{Guzman and Maalej(2014)}]{guzman2014sentiment}
\bibinfo{author}{Guzman, E.}, \bibinfo{author}{Maalej, W.}, \bibinfo{year}{2014}.
\newblock \bibinfo{title}{Sentiment analysis of mobile app reviews}, in: \bibinfo{booktitle}{2014 IEEE 22nd International Requirements Engineering Conference (RE)}, \bibinfo{organization}{IEEE}. pp. \bibinfo{pages}{383--388}.
\bibitem[{Harbers et~al.(2015)Harbers, Detweiler and Neerincx}]{harbers2015embedding}
\bibinfo{author}{Harbers, M.}, \bibinfo{author}{Detweiler, C.}, \bibinfo{author}{Neerincx, M.A.}, \bibinfo{year}{2015}.
\newblock \bibinfo{title}{Embedding stakeholder values in the requirements engineering process}, in: \bibinfo{booktitle}{International working conference on requirements engineering: Foundation for software quality}, \bibinfo{organization}{Springer}. pp. \bibinfo{pages}{318--332}.
\bibitem[{Harrington et~al.(2019)Harrington, Erete and Piper}]{harrington2019deconstructing}
\bibinfo{author}{Harrington, C.N.}, \bibinfo{author}{Erete, S.}, \bibinfo{author}{Piper, A.M.}, \bibinfo{year}{2019}.
\newblock \bibinfo{title}{Deconstructing community-based collaborative design: Towards more equitable participatory design engagements}, in: \bibinfo{booktitle}{Proceedings of the ACM on Human-Computer Interaction}, pp. \bibinfo{pages}{1--25}.
\newblock \DOIprefix\doi{10.1145/3359318}.
\bibitem[{Hidellaarachchi et~al.(2021)Hidellaarachchi, Grundy, Hoda and Madampe}]{hidellaarachchi2021effects}
\bibinfo{author}{Hidellaarachchi, D.}, \bibinfo{author}{Grundy, J.}, \bibinfo{author}{Hoda, R.}, \bibinfo{author}{Madampe, K.}, \bibinfo{year}{2021}.
\newblock \bibinfo{title}{The effects of human aspects on the requirements engineering process: A systematic literature review}.
\newblock \bibinfo{journal}{IEEE Transactions on Software Engineering} \bibinfo{volume}{48}, \bibinfo{pages}{2105--2127}.
\bibitem[{Hussain et~al.(2022)Hussain, Shahin, Hoda, Whittle, Perera, Nurwidyantoro, Shams and Oliver}]{hussain2022can}
\bibinfo{author}{Hussain, W.}, \bibinfo{author}{Shahin, M.}, \bibinfo{author}{Hoda, R.}, \bibinfo{author}{Whittle, J.}, \bibinfo{author}{Perera, H.}, \bibinfo{author}{Nurwidyantoro, A.}, \bibinfo{author}{Shams, R.A.}, \bibinfo{author}{Oliver, G.}, \bibinfo{year}{2022}.
\newblock \bibinfo{title}{How can human values be addressed in agile methods? a case study on safe}.
\newblock \bibinfo{journal}{IEEE Transactions on Software Engineering} \bibinfo{volume}{48}, \bibinfo{pages}{5158--5175}.
\bibitem[{{InfoWorld}(2025)}]{infoworld2025genai}
\bibinfo{author}{{InfoWorld}}, \bibinfo{year}{2025}.
\newblock \bibinfo{title}{How to use genai for requirements gathering and agile user stories}.
\newblock \URLprefix \url{https://www.infoworld.com/article/3980319/how-to-use-genai-for-requirements-gathering-and-agile-user-stories.html}.
\bibitem[{Jagosh et~al.(2015)Jagosh, Bush, Salsberg, Macaulay, Greenhalgh, Wong, Cargo, Green, Herbert and Pluye}]{jagosh2015realist}
\bibinfo{author}{Jagosh, J.}, \bibinfo{author}{Bush, P.L.}, \bibinfo{author}{Salsberg, J.}, \bibinfo{author}{Macaulay, A.C.}, \bibinfo{author}{Greenhalgh, T.}, \bibinfo{author}{Wong, G.}, \bibinfo{author}{Cargo, M.}, \bibinfo{author}{Green, L.W.}, \bibinfo{author}{Herbert, C.P.}, \bibinfo{author}{Pluye, P.}, \bibinfo{year}{2015}.
\newblock \bibinfo{title}{A realist evaluation of community-based participatory research: Partnership synergy, trust building and related ripple effects}.
\newblock \bibinfo{journal}{BMC Public Health} \bibinfo{volume}{15}, \bibinfo{pages}{725}.
\newblock \DOIprefix\doi{10.1186/s12889-015-1949-1}.
\bibitem[{Karkee and Comfort(2016)}]{Karkee2016Nepal}
\bibinfo{author}{Karkee, R.}, \bibinfo{author}{Comfort, J.}, \bibinfo{year}{2016}.
\newblock \bibinfo{title}{Ngos, foreign aid, and development in nepal}.
\newblock \bibinfo{journal}{Frontiers in Public Health} \bibinfo{volume}{4}.
\newblock \DOIprefix\doi{10.3389/fpubh.2016.00177}.
\bibitem[{Kathayat(2024)}]{kathayat2024investigating}
\bibinfo{author}{Kathayat, B.B.}, \bibinfo{year}{2024}.
\newblock \bibinfo{title}{Investigating public trust and ethical leadership in nepalese cooperatives}.
\newblock \bibinfo{journal}{Journal of Nepalese Management and Research} \bibinfo{volume}{6}, \bibinfo{pages}{17--28}.
\bibitem[{Kerzner et~al.(2018)Kerzner, Goodwin, Dykes, Jones and Meyer}]{kerzner2018framework}
\bibinfo{author}{Kerzner, E.}, \bibinfo{author}{Goodwin, S.}, \bibinfo{author}{Dykes, J.}, \bibinfo{author}{Jones, S.}, \bibinfo{author}{Meyer, M.}, \bibinfo{year}{2018}.
\newblock \bibinfo{title}{A framework for creative visualization-opportunities workshops}.
\newblock \bibinfo{journal}{IEEE transactions on visualization and computer graphics} \bibinfo{volume}{25}, \bibinfo{pages}{748--758}.
\bibitem[{Kim et~al.(2021)Kim, Humble, Debois, Willis and Forsgren}]{kim2021devops}
\bibinfo{author}{Kim, G.}, \bibinfo{author}{Humble, J.}, \bibinfo{author}{Debois, P.}, \bibinfo{author}{Willis, J.}, \bibinfo{author}{Forsgren, N.}, \bibinfo{year}{2021}.
\newblock \bibinfo{title}{The DevOps handbook: How to create world-class agility, reliability, \& security in technology organizations}.
\newblock \bibinfo{publisher}{It Revolution}.
\bibitem[{Kim and Hwang(2026)}]{kim2026forward}
\bibinfo{author}{Kim, J.}, \bibinfo{author}{Hwang, H.}, \bibinfo{year}{2026}.
\newblock \bibinfo{title}{Forward deployed engineering: A taxonomy and definition}.
\newblock \bibinfo{journal}{Available at SSRN 6374660} .
\bibitem[{Kujala(2003)}]{kujala2003user}
\bibinfo{author}{Kujala, S.}, \bibinfo{year}{2003}.
\newblock \bibinfo{title}{User involvement: a review of the benefits and challenges}.
\newblock \bibinfo{journal}{Behaviour \& information technology} \bibinfo{volume}{22}, \bibinfo{pages}{1--16}.
\bibitem[{Kujala et~al.(2005)Kujala, Kauppinen, Lehtola and Kojo}]{kujala2005role}
\bibinfo{author}{Kujala, S.}, \bibinfo{author}{Kauppinen, M.}, \bibinfo{author}{Lehtola, L.}, \bibinfo{author}{Kojo, T.}, \bibinfo{year}{2005}.
\newblock \bibinfo{title}{The role of user involvement in requirements quality and project success}, in: \bibinfo{booktitle}{13th IEEE International Conference on Requirements Engineering (RE'05)}, \bibinfo{organization}{IEEE}. pp. \bibinfo{pages}{75--84}.
\bibitem[{Kurtanović and Maalej(2017)}]{kurtanovic2017requirements}
\bibinfo{author}{Kurtanović, Z.}, \bibinfo{author}{Maalej, W.}, \bibinfo{year}{2017}.
\newblock \bibinfo{title}{Requirements classification of app reviews}, in: \bibinfo{booktitle}{2017 IEEE 25th International Requirements Engineering Conference Workshops (REW)}, \bibinfo{organization}{IEEE}. pp. \bibinfo{pages}{208--213}.
\bibitem[{Le~Dantec and Fox(2015)}]{le2015strangers}
\bibinfo{author}{Le~Dantec, C.A.}, \bibinfo{author}{Fox, S.}, \bibinfo{year}{2015}.
\newblock \bibinfo{title}{Strangers at the gate: Gaining access, building rapport, and co-constructing community-based research}, in: \bibinfo{booktitle}{Proceedings of the 18th ACM conference on computer supported cooperative work \& social computing}, pp. \bibinfo{pages}{1348--1358}.
\bibitem[{Levina(2024)}]{levina2024incorporating}
\bibinfo{author}{Levina, O.}, \bibinfo{year}{2024}.
\newblock \bibinfo{title}{Incorporating ethical aspects in information systems requirements engineering}, in: \bibinfo{booktitle}{International Conference on Business Informatics Research}, \bibinfo{organization}{Springer}. pp. \bibinfo{pages}{153--161}.
\bibitem[{Liu(2025)}]{Liu2025HumanAICoCreation}
\bibinfo{author}{Liu, Z.}, \bibinfo{year}{2025}.
\newblock \bibinfo{title}{Human-ai co-creation: A framework for collaborative design in intelligent systems}, in: \bibinfo{editor}{Ahram, T.}, \bibinfo{editor}{Karwowski, W.}, \bibinfo{editor}{Kalra, J.} (Eds.), \bibinfo{booktitle}{Human Factors in Design, Engineering, and Computing}, \bibinfo{publisher}{AHFE International}, \bibinfo{address}{USA}. pp. \bibinfo{pages}{2244--2252}.
\newblock \URLprefix \url{https://doi.org/10.54941/ahfe1007036}, \DOIprefix\doi{10.54941/ahfe1007036}.
\bibitem[{Maalej et~al.(2025)Maalej, Biryuk, Wei and Panse}]{maalej2025automated}
\bibinfo{author}{Maalej, W.}, \bibinfo{author}{Biryuk, V.}, \bibinfo{author}{Wei, J.}, \bibinfo{author}{Panse, F.}, \bibinfo{year}{2025}.
\newblock \bibinfo{title}{On the automated processing of user feedback}, in: \bibinfo{booktitle}{Handbook on Natural Language Processing for Requirements Engineering}. \bibinfo{publisher}{Springer}, pp. \bibinfo{pages}{279--308}.
\bibitem[{Macaulay(2012)}]{macaulay2012requirements}
\bibinfo{author}{Macaulay, L.A.}, \bibinfo{year}{2012}.
\newblock \bibinfo{title}{Requirements engineering}.
\newblock \bibinfo{publisher}{Springer Science \& Business Media}.
\bibitem[{Mannheim et~al.(2019)Mannheim, Schwartz, Xi, Buttigieg, McDonnell-Naughton, Wouters and Van~Zaalen}]{mannheim2019inclusion}
\bibinfo{author}{Mannheim, I.}, \bibinfo{author}{Schwartz, E.}, \bibinfo{author}{Xi, W.}, \bibinfo{author}{Buttigieg, S.C.}, \bibinfo{author}{McDonnell-Naughton, M.}, \bibinfo{author}{Wouters, E.J.}, \bibinfo{author}{Van~Zaalen, Y.}, \bibinfo{year}{2019}.
\newblock \bibinfo{title}{Inclusion of older adults in the research and design of digital technology}.
\newblock \bibinfo{journal}{International journal of environmental research and public health} \bibinfo{volume}{16}, \bibinfo{pages}{3718}.
\bibitem[{Marques et~al.(2024)Marques, Silva and Bernardino}]{marques2024using}
\bibinfo{author}{Marques, N.}, \bibinfo{author}{Silva, R.R.}, \bibinfo{author}{Bernardino, J.}, \bibinfo{year}{2024}.
\newblock \bibinfo{title}{Using chatgpt in software requirements engineering: A comprehensive review}.
\newblock \bibinfo{journal}{Future Internet} \bibinfo{volume}{16}, \bibinfo{pages}{180}.
\bibitem[{McKim(2023)}]{mckim2023meaningful}
\bibinfo{author}{McKim, C.}, \bibinfo{year}{2023}.
\newblock \bibinfo{title}{Meaningful member-checking: A structured approach to member-checking}.
\newblock \bibinfo{journal}{American Journal of Qualitative Research} \bibinfo{volume}{7}, \bibinfo{pages}{41--52}.
\bibitem[{Mitchell et~al.(2025)Mitchell, Ghosh, Luccioni and Pistilli}]{mitchell2025fully}
\bibinfo{author}{Mitchell, M.}, \bibinfo{author}{Ghosh, A.}, \bibinfo{author}{Luccioni, A.S.}, \bibinfo{author}{Pistilli, G.}, \bibinfo{year}{2025}.
\newblock \bibinfo{title}{Fully autonomous ai agents should not be developed}.
\newblock \bibinfo{journal}{arXiv preprint arXiv:2502.02649} .
\bibitem[{Mougouei et~al.(2018)Mougouei, Perera, Hussain, Shams and Whittle}]{mougouei2018operationalizing}
\bibinfo{author}{Mougouei, D.}, \bibinfo{author}{Perera, H.}, \bibinfo{author}{Hussain, W.}, \bibinfo{author}{Shams, R.}, \bibinfo{author}{Whittle, J.}, \bibinfo{year}{2018}.
\newblock \bibinfo{title}{Operationalizing human values in software: A research roadmap}, in: \bibinfo{booktitle}{Proceedings of the 2018 26th ACM joint meeting on european software engineering conference and symposium on the foundations of software engineering}, pp. \bibinfo{pages}{780--784}.
\bibitem[{Mukherjee and Chang(2025)}]{mukherjee2025agentic}
\bibinfo{author}{Mukherjee, A.}, \bibinfo{author}{Chang, H.H.}, \bibinfo{year}{2025}.
\newblock \bibinfo{title}{Agentic ai: Autonomy, accountability, and the algorithmic society}.
\newblock \bibinfo{journal}{arXiv preprint arXiv:2502.00289} .
\bibitem[{Noble(2018)}]{noble2018algorithms}
\bibinfo{author}{Noble, S.U.}, \bibinfo{year}{2018}.
\newblock \bibinfo{title}{Algorithms of oppression: How search engines reinforce racism}, in: \bibinfo{booktitle}{Algorithms of oppression}. \bibinfo{publisher}{New York university press}.
\bibitem[{Nuseibeh(2025)}]{nuseibeh2025engineering}
\bibinfo{author}{Nuseibeh, B.}, \bibinfo{year}{2025}.
\newblock \bibinfo{title}{Engineering within boundaries when software has none}.
\newblock \bibinfo{journal}{IEEE Transactions on Software Engineering} \bibinfo{volume}{51}, \bibinfo{pages}{677--680}.
\newblock \URLprefix \url{https://doi.org/10.1109/TSE.2025.3541189}, \DOIprefix\doi{10.1109/TSE.2025.3541189}.
\bibitem[{Panday et~al.(2017)Panday, Bissell, Teijlingen and Simkhada}]{Panday2017Health}
\bibinfo{author}{Panday, S.}, \bibinfo{author}{Bissell, P.}, \bibinfo{author}{Teijlingen, E.v.}, \bibinfo{author}{Simkhada, P.}, \bibinfo{year}{2017}.
\newblock \bibinfo{title}{The contribution of female community health volunteers (fchvs) to maternity care in nepal: a qualitative study}.
\newblock \bibinfo{journal}{BMC Health Services Research} \bibinfo{volume}{17}.
\newblock \DOIprefix\doi{10.1186/s12913-017-2567-7}.
\bibitem[{on~Patient~Safety and Technology(2012)}]{committee2012health}
\bibinfo{author}{on~Patient~Safety, C.}, \bibinfo{author}{Technology, H.I.}, \bibinfo{year}{2012}.
\newblock \bibinfo{title}{Health it and patient safety: building safer systems for better care} .
\bibitem[{Perera et~al.(2021)Perera, Hoda, Shams, Nurwidyantoro, Shahin, Hussain and Whittle}]{perera2021impact}
\bibinfo{author}{Perera, H.}, \bibinfo{author}{Hoda, R.}, \bibinfo{author}{Shams, R.A.}, \bibinfo{author}{Nurwidyantoro, A.}, \bibinfo{author}{Shahin, M.}, \bibinfo{author}{Hussain, W.}, \bibinfo{author}{Whittle, J.}, \bibinfo{year}{2021}.
\newblock \bibinfo{title}{The impact of considering human values during requirements engineering activities}.
\newblock \bibinfo{journal}{arXiv preprint arXiv:2111.15293} .
\bibitem[{Potts and Newstetter(1997)}]{potts1997naturalistic}
\bibinfo{author}{Potts, C.}, \bibinfo{author}{Newstetter, W.C.}, \bibinfo{year}{1997}.
\newblock \bibinfo{title}{Naturalistic inquiry and requirements engineering: reconciling their theoretical foundations}, in: \bibinfo{booktitle}{Proceedings of ISRE'97: 3rd IEEE International Symposium on Requirements Engineering}, \bibinfo{organization}{IEEE}. pp. \bibinfo{pages}{118--127}.
\bibitem[{Robinson et~al.(2025)Robinson, Cabrera, Gordon, Lawrence and Mennen}]{robinson2025requirements}
\bibinfo{author}{Robinson, D.}, \bibinfo{author}{Cabrera, C.}, \bibinfo{author}{Gordon, A.D.}, \bibinfo{author}{Lawrence, N.D.}, \bibinfo{author}{Mennen, L.}, \bibinfo{year}{2025}.
\newblock \bibinfo{title}{Requirements are all you need: The final frontier for end-user software engineering}.
\newblock \bibinfo{journal}{ACM Transactions on Software Engineering and Methodology} \bibinfo{volume}{34}, \bibinfo{pages}{1--22}.
\newblock \DOIprefix\doi{10.1145/3708524}.
\bibitem[{Ronanki et~al.(2023)Ronanki, Berger and Horkoff}]{ronanki2023chatgpt}
\bibinfo{author}{Ronanki, K.}, \bibinfo{author}{Berger, C.}, \bibinfo{author}{Horkoff, J.}, \bibinfo{year}{2023}.
\newblock \bibinfo{title}{Investigating chatgpt's potential to assist in requirements elicitation processes}.
\newblock \bibinfo{journal}{arXiv preprint arXiv:2307.07381} \URLprefix \url{https://arxiv.org/abs/2307.07381}.
\bibitem[{Ryan and Deci(2000)}]{ryan2000self}
\bibinfo{author}{Ryan, R.M.}, \bibinfo{author}{Deci, E.L.}, \bibinfo{year}{2000}.
\newblock \bibinfo{title}{Self-determination theory and the facilitation of intrinsic motivation, social development, and well-being.}
\newblock \bibinfo{journal}{American psychologist} \bibinfo{volume}{55}, \bibinfo{pages}{68}.
\bibitem[{Sami et~al.(2024)Sami, Waseem, Zhang, Rasheed, Systä and Abrahamsson}]{sami2024ai}
\bibinfo{author}{Sami, M.A.}, \bibinfo{author}{Waseem, M.}, \bibinfo{author}{Zhang, Z.}, \bibinfo{author}{Rasheed, Z.}, \bibinfo{author}{Systä, K.}, \bibinfo{author}{Abrahamsson, P.}, \bibinfo{year}{2024}.
\newblock \bibinfo{title}{Ai-based multiagent approach for requirements elicitation and analysis}.
\newblock \bibinfo{journal}{arXiv preprint arXiv:2409.00038} \URLprefix \url{https://arxiv.org/abs/2409.00038}.
\bibitem[{Sch{\"o}n(2017)}]{schon2017reflective}
\bibinfo{author}{Sch{\"o}n, D.A.}, \bibinfo{year}{2017}.
\newblock \bibinfo{title}{The reflective practitioner: How professionals think in action}.
\newblock \bibinfo{publisher}{Routledge}.
\bibitem[{Schwartz(2012)}]{schwartz2012overview}
\bibinfo{author}{Schwartz, S.H.}, \bibinfo{year}{2012}.
\newblock \bibinfo{title}{An overview of the schwartz theory of basic values}.
\newblock \bibinfo{journal}{Online readings in Psychology and Culture} \bibinfo{volume}{2}, \bibinfo{pages}{11}.
\bibitem[{of~Sciences et~al.(2020)of~Sciences, Medicine, Board, on~Engineering, Sciences, on~Energy, Systems, on~Assessment~of Technologies, for Reducing the Fuel Consumption~of Medium, Vehicles et~al.}]{national2020reducing}
\bibinfo{author}{of~Sciences, N.A.}, \bibinfo{author}{Medicine}, \bibinfo{author}{Board, T.R.}, \bibinfo{author}{on~Engineering, D.}, \bibinfo{author}{Sciences, P.}, \bibinfo{author}{on~Energy, B.}, \bibinfo{author}{Systems, E.}, \bibinfo{author}{on~Assessment~of Technologies, C.}, \bibinfo{author}{for Reducing the Fuel Consumption~of Medium, A.}, \bibinfo{author}{Vehicles, H.D.}, et~al., \bibinfo{year}{2020}.
\newblock \bibinfo{title}{Reducing fuel consumption and greenhouse gas emissions of medium-and heavy-duty vehicles, phase two}.
\newblock \bibinfo{publisher}{National Academies Press}.
\bibitem[{Segura and Barbosa(2013)}]{segura2013uiskei++}
\bibinfo{author}{Segura, V.C.}, \bibinfo{author}{Barbosa, S.D.}, \bibinfo{year}{2013}.
\newblock \bibinfo{title}{Uiskei++ multi-device wizard of oz prototyping}, in: \bibinfo{booktitle}{Proceedings of the 5th ACM SIGCHI symposium on Engineering interactive computing systems}, pp. \bibinfo{pages}{171--174}.
\bibitem[{Shahin et~al.(2022)Shahin, Hussain, Nurwidyantoro, Perera, Shams, Grundy and Whittle}]{Shahin_2022}
\bibinfo{author}{Shahin, M.}, \bibinfo{author}{Hussain, W.}, \bibinfo{author}{Nurwidyantoro, A.}, \bibinfo{author}{Perera, H.}, \bibinfo{author}{Shams, R.}, \bibinfo{author}{Grundy, J.}, \bibinfo{author}{Whittle, J.}, \bibinfo{year}{2022}.
\newblock \bibinfo{title}{Operationalizing human values in software engineering: A survey}.
\newblock \bibinfo{journal}{IEEE Access} \bibinfo{volume}{10}, \bibinfo{pages}{75269–75295}.
\newblock \URLprefix \url{http://dx.doi.org/10.1109/ACCESS.2022.3190975}, \DOIprefix\doi{10.1109/access.2022.3190975}.
\bibitem[{Shinn(2007)}]{shinn2007international}
\bibinfo{author}{Shinn, M.}, \bibinfo{year}{2007}.
\newblock \bibinfo{title}{International homelessness: Policy, socio-cultural, and individual perspectives}.
\newblock \bibinfo{journal}{Journal of Social Issues} \bibinfo{volume}{63}, \bibinfo{pages}{657--677}.
\bibitem[{Simonsen and Robertson(2013)}]{simonsen2013routledge}
\bibinfo{author}{Simonsen, J.}, \bibinfo{author}{Robertson, T.}, \bibinfo{year}{2013}.
\newblock \bibinfo{title}{Routledge international handbook of participatory design}. volume \bibinfo{volume}{711}.
\newblock \bibinfo{publisher}{Routledge New York}.
\bibitem[{Smith et~al.(2018)Smith, Bossen and Kanstrup}]{smith2018pd}
\bibinfo{author}{Smith, R.C.}, \bibinfo{author}{Bossen, C.}, \bibinfo{author}{Kanstrup, A.M.}, \bibinfo{year}{2018}.
\newblock \bibinfo{title}{Participatory design for sustainable social change}.
\newblock \bibinfo{journal}{Design Studies} \bibinfo{volume}{59}, \bibinfo{pages}{11--36}.
\newblock \DOIprefix\doi{10.1016/j.destud.2018.05.005}.
\bibitem[{Sommerville and Sawyer(1997)}]{sommerville1997requirements}
\bibinfo{author}{Sommerville, I.}, \bibinfo{author}{Sawyer, P.}, \bibinfo{year}{1997}.
\newblock \bibinfo{title}{Requirements engineering: a good practice guide}.
\newblock \bibinfo{publisher}{John Wiley \& Sons, Inc.}
\bibitem[{Moises~de Souza et~al.(2023)Moises~de Souza, Soares~Cruzes, Jaccheri and Krogstie}]{moises2023social}
\bibinfo{author}{Moises~de Souza, A.C.}, \bibinfo{author}{Soares~Cruzes, D.}, \bibinfo{author}{Jaccheri, L.}, \bibinfo{author}{Krogstie, J.}, \bibinfo{year}{2023}.
\newblock \bibinfo{title}{Social sustainability approaches for software development: A systematic literature review}, in: \bibinfo{booktitle}{International Conference on Product-Focused Software Process Improvement}, \bibinfo{organization}{Springer}. pp. \bibinfo{pages}{478--494}.
\bibitem[{Subedi and Bhandari(2025)}]{subedi2025ethical}
\bibinfo{author}{Subedi, D.P.}, \bibinfo{author}{Bhandari, D.R.}, \bibinfo{year}{2025}.
\newblock \bibinfo{title}{Ethical business practices and employee retention in nepal: Investigating the mediating role of employee loyalty}.
\newblock \bibinfo{journal}{Victoria Journal of Management} \bibinfo{volume}{1}, \bibinfo{pages}{51--75}.
\bibitem[{Sutcliffe(2002)}]{sutcliffe2002user}
\bibinfo{author}{Sutcliffe, A.}, \bibinfo{year}{2002}.
\newblock \bibinfo{title}{User-centred requirements engineering}.
\newblock \bibinfo{publisher}{Springer Science \& Business Media}.
\bibitem[{Sutcliffe et~al.(2005)Sutcliffe, Fickas and Sohlberg}]{sutcliffe2005personal}
\bibinfo{author}{Sutcliffe, A.}, \bibinfo{author}{Fickas, S.}, \bibinfo{author}{Sohlberg, M.M.}, \bibinfo{year}{2005}.
\newblock \bibinfo{title}{Personal and contextual requirements engineering}, in: \bibinfo{booktitle}{13th IEEE International Conference on Requirements Engineering (RE'05)}, \bibinfo{organization}{IEEE}. pp. \bibinfo{pages}{19--28}.
\bibitem[{Thew and Sutcliffe(2018)}]{thew2018value}
\bibinfo{author}{Thew, S.}, \bibinfo{author}{Sutcliffe, A.}, \bibinfo{year}{2018}.
\newblock \bibinfo{title}{Value-based requirements engineering: method and experience}.
\newblock \bibinfo{journal}{Requirements engineering} \bibinfo{volume}{23}, \bibinfo{pages}{443--464}.
\bibitem[{Thompson(2021)}]{thompson2021reflective}
\bibinfo{author}{Thompson, C.}, \bibinfo{year}{2021}.
\newblock \bibinfo{title}{Reflective practice for professional development: A guide for teachers}.
\newblock \bibinfo{publisher}{Routledge}.
\bibitem[{Tizard et~al.(2024)Tizard, Rietz and Blincoe}]{tizard2024elicitation}
\bibinfo{author}{Tizard, J.}, \bibinfo{author}{Rietz, T.}, \bibinfo{author}{Blincoe, K.}, \bibinfo{year}{2024}.
\newblock \bibinfo{title}{Elicitation revisited for more inclusive requirements engineering}, in: \bibinfo{booktitle}{Equity, Diversity, and Inclusion in Software Engineering: Best Practices and Insights}. \bibinfo{publisher}{Apress Berkeley, CA}, pp. \bibinfo{pages}{91--104}.
\bibitem[{Tizard et~al.(2022)Tizard, Rietz, Liu and Blincoe}]{TizardRLB22}
\bibinfo{author}{Tizard, J.}, \bibinfo{author}{Rietz, T.}, \bibinfo{author}{Liu, X.}, \bibinfo{author}{Blincoe, K.}, \bibinfo{year}{2022}.
\newblock \bibinfo{title}{Voice of the users: An extended study of software feedback engagement}.
\newblock \bibinfo{journal}{Requirements Engineering} \bibinfo{volume}{27}, \bibinfo{pages}{293--315}.
\newblock \DOIprefix\doi{10.1007/s00766-021-00357-1}.
\bibitem[{Tronnier and et~al.(2024)}]{tronnier2024bias}
\bibinfo{author}{Tronnier, K.}, \bibinfo{author}{et~al.}, \bibinfo{year}{2024}.
\newblock \bibinfo{title}{Mind the gap: Bias in generative ai}.
\newblock \bibinfo{howpublished}{\url{https://www.equalvoice.ch/wp-content/uploads/2025/01/IMD_Mind-The-Gap_Bias-In-AI_Whitepaper_DIGITAL.pdf}}.
\bibitem[{{UNDP}(2024)}]{undp2024gendersensitive}
\bibinfo{author}{{UNDP}}, \bibinfo{year}{2024}.
\newblock \bibinfo{title}{Understanding the implications of genai on gender}.
\newblock \bibinfo{howpublished}{\url{https://www.undp.org/sites/g/files/zskgke326/files/2024-08/report_understanding_the_implications_of_genai_on_gender_undp_aapti_.pdf}}.
\bibitem[{Unertl et~al.(2015)Unertl, Schaefbauer, Campbell, Senteio, Siek, Krupinski, Gibbons and Grando}]{unertl2015cbpr}
\bibinfo{author}{Unertl, K.M.}, \bibinfo{author}{Schaefbauer, C.L.}, \bibinfo{author}{Campbell, T.R.}, \bibinfo{author}{Senteio, C.}, \bibinfo{author}{Siek, K.A.}, \bibinfo{author}{Krupinski, E.A.}, \bibinfo{author}{Gibbons, M.C.}, \bibinfo{author}{Grando, M.A.}, \bibinfo{year}{2015}.
\newblock \bibinfo{title}{Integrating community-based participatory research and informatics approaches to improve the engagement and health of underserved populations}.
\newblock \bibinfo{journal}{Journal of the American Medical Informatics Association} \bibinfo{volume}{22}, \bibinfo{pages}{471--480}.
\newblock \DOIprefix\doi{10.1136/amiajnl-2014-003161}.
\bibitem[{Wallerstein and Duran(2006)}]{wallerstein2006cbpr}
\bibinfo{author}{Wallerstein, N.}, \bibinfo{author}{Duran, B.}, \bibinfo{year}{2006}.
\newblock \bibinfo{title}{Using community-based participatory research to address health disparities}.
\newblock \bibinfo{journal}{Health Promotion Practice} \bibinfo{volume}{7}, \bibinfo{pages}{312--323}.
\newblock \DOIprefix\doi{10.1177/1524839906289376}.
\bibitem[{Wallerstein and Duran(2010)}]{wallerstein2010cbpr}
\bibinfo{author}{Wallerstein, N.}, \bibinfo{author}{Duran, B.}, \bibinfo{year}{2010}.
\newblock \bibinfo{title}{Community-based participatory research contributions to intervention research: The intersection of science and practice to improve health equity}.
\newblock \bibinfo{journal}{American Journal of Public Health} \bibinfo{volume}{100}, \bibinfo{pages}{S40--S46}.
\newblock \DOIprefix\doi{10.2105/AJPH.2009.184036}.
\bibitem[{Wei(2024)}]{wei2024requirements}
\bibinfo{author}{Wei, B.}, \bibinfo{year}{2024}.
\newblock \bibinfo{title}{Requirements are all you need: From requirements to code with llms}, in: \bibinfo{booktitle}{2024 IEEE 32nd International Requirements Engineering Conference (RE)}, \bibinfo{organization}{IEEE}. pp. \bibinfo{pages}{416--422}.
\bibitem[{Werthner(2020)}]{werthner2020vienna}
\bibinfo{author}{Werthner, H.}, \bibinfo{year}{2020}.
\newblock \bibinfo{title}{The vienna manifesto on digital humanism}, in: \bibinfo{booktitle}{Digital transformation and ethics}, \bibinfo{organization}{Ecowin}. pp. \bibinfo{pages}{338--357}.
\bibitem[{Whittle and et~al.(2014)}]{whittle2014citizen}
\bibinfo{author}{Whittle, J.}, \bibinfo{author}{et~al.}, \bibinfo{year}{2014}.
\newblock \bibinfo{title}{Citizen-centered design: A catalyst for innovation}, in: \bibinfo{booktitle}{Proceedings of the 2014 ACM Conference on Designing Interactive Systems}, \bibinfo{publisher}{ACM}, \bibinfo{address}{New York, NY}. pp. \bibinfo{pages}{1--10}.
\newblock \DOIprefix\doi{10.1145/2598510.2598511}.
\bibitem[{Whittle et~al.(2019a)Whittle, Ferrario, Simm and Hussain}]{whittle2019case}
\bibinfo{author}{Whittle, J.}, \bibinfo{author}{Ferrario, M.A.}, \bibinfo{author}{Simm, W.}, \bibinfo{author}{Hussain, W.}, \bibinfo{year}{2019}a.
\newblock \bibinfo{title}{A case for human values in software engineering}.
\newblock \bibinfo{journal}{IEEE software} \bibinfo{volume}{38}, \bibinfo{pages}{106--113}.
\bibitem[{Whittle et~al.(2019b)Whittle, Ferrario et~al.}]{whittle2019values}
\bibinfo{author}{Whittle, J.}, \bibinfo{author}{Ferrario, M.A.}, et~al., \bibinfo{year}{2019}b.
\newblock \bibinfo{title}{Is your software valueless? how integrating human values into software engineering can transform society}.
\newblock \bibinfo{journal}{IEEE Software} \bibinfo{volume}{36}, \bibinfo{pages}{48--55}.
\newblock \DOIprefix\doi{10.1109/MS.2019.2927861}.
\bibitem[{Wilson et~al.(2019)Wilson, Hoffman and Morgenstern}]{wilson2019predictive}
\bibinfo{author}{Wilson, B.}, \bibinfo{author}{Hoffman, J.}, \bibinfo{author}{Morgenstern, J.}, \bibinfo{year}{2019}.
\newblock \bibinfo{title}{Predictive inequity in object detection}.
\newblock \bibinfo{journal}{arXiv preprint arXiv:1902.11097} .
\bibitem[{Wilson-Raybould(2024)}]{wilson-raybould2024true}
\bibinfo{author}{Wilson-Raybould, J.}, \bibinfo{year}{2024}.
\newblock \bibinfo{title}{True Reconciliation: How to Be a Force for Change}.
\newblock \bibinfo{publisher}{McClelland \& Stewart}, \bibinfo{address}{Toronto, Canada}.
\bibitem[{Zou et~al.(2025)Zou, Kuek, Feng and Cheng}]{zou2025digital}
\bibinfo{author}{Zou, Y.}, \bibinfo{author}{Kuek, F.}, \bibinfo{author}{Feng, W.}, \bibinfo{author}{Cheng, X.}, \bibinfo{year}{2025}.
\newblock \bibinfo{title}{Digital learning in the 21st century: trends, challenges, and innovations in technology integration}, in: \bibinfo{booktitle}{Frontiers in Education}, \bibinfo{organization}{Frontiers Media SA}. p. \bibinfo{pages}{1562391}.

\end{thebibliography}




\end{document}